\documentclass{article}
\usepackage[dvips]{rotating}

\title{Microwave Background Anisotropy and Large Scale Structure in Warm Dark Matter Models}
\author{Stephen D. Burns\thanks{e-mail address:  burns@physics.ucla.edu}
 \\ \begin{em}Department of Physics and Astronomy, University of California\end{em} \\ \begin{em}Los Angeles, California 90025\end{em}}

\begin{document}

\maketitle

\begin{abstract}
Large scale structure and microwave background anisotropies are studied for warm dark matter models.  Two warm dark matter candidates are considered:  gravitinos and sterile neutrinos.  Linear large scale structure properties such as $\sigma_{8}$ and the excess power are calculated, as well as microwave background anisotropies.  A rather robust feature of warm dark matter models is that the large scale structure properties are similar to those of mixed dark matter, but that the microwave background anisotropy is very similar to that of standard cold dark matter.
\end{abstract}

\section{Introduction:  CDM, HDM and WDM}
  
One of the outstanding problems in cosmology today is that of dark matter and its role in structure formation.  Dark matter is usually classified by its velocity dispersion at matter-radiation equality, when structure formation commences.  At that time, cold dark matter has a negligible mean velocity, which preferentially leads to the collapse of small scale objects.  Hot dark matter on the other hand, has a very high mean velocity at matter-radiation equality, which means that the hot dark matter is able to free-stream out of the gravitational potentials.  The typical hot dark matter candidate is a relatively light neutrino.  For a neutrino with a mass $m_{\nu}$, the free-streaming length is \cite{KT}
\begin{equation}
 \lambda_{\nu FS}\approx 71.4\left(\frac{m_{\nu}}{10\mbox{eV}}\right)^{-1}\left(\frac{10.75}{g_{\ast}(T_{D})} \right)^{1/3} \mbox{Mpc},
\end{equation}
where $g_{\ast}(T_{D})$ is the number of relativistic degrees of freedom when the particle decouples from the plasma.  Neutrinos, for example, decouple when $g_{\ast}(T_{D})$=10.75.  This free-streaming of power to larger scales means that larger objects tend to form first, later fragmenting into smaller objects.

These are the simplest models of structure formation.  In both of these scenarios, after fixing the initial power spectrum and the classical cosmological parameters $\Omega$, the cosmological constant $\Lambda$, the baryon density $\Omega_{b}$, and the Hubble constant $H_{0}$, the model is completely specified.  Typically, for example, $P(k)=k$, $\Omega=1$, $\Lambda=0$, $0.01\leq\Omega_{b}\leq0.1$, and $0.5\leq h\leq1$.  It has long been recognized, however, that both of these models have serious difficulties.  Hot dark matter, for example, tends to form structures at very low redshifts ($z\approx1$) whereas many galaxies and quasars are observed at redshifts up to $z$ of 4.  Cold dark matter, on the other hand, has no difficulty forming galaxies, but tends to produce too much power on small scales and not enough on large scales.

Previous work on solving this problem has tended to focus on three solutions:  mixed dark matter \cite{ShafiMDM,DavisMDM,TaylorMDM,vanDalenMDM,KlypinMDM}, a tilted power spectrum \cite{tiltedCDM}, and a cosmological constant \cite{Peebleslambda,Turnerlambda1,Efstathioulambda,Turnerlambda2}.  Mixed dark matter models include some hot dark matter to complement the cold dark matter.  The free-streaming neutrinos can then disperse some of the small scale power to larger scales.  The second approach is a tilt in the primordial power spectrum, $P(k)=k^{n}$, with $n<1$.  This is a rather generic prediction of most inflationary models.  Since the inflationary potential is not constant during the inflationary epoch, the spectrum of fluctuations is not exactly scale invariant.  This tilt in the power spectrum results in less power at small scales (high $k$).  Finally, there is the cosmological constant.  For a fixed $\Omega$, adding a cosmological constant at the expense of the matter both decreases the matter density, and, through spatial repulsion, disperses power to larger scales.  

In this article, we consider another solution to this problem, warm dark matter.  Warm dark matter is an alternative to mixed dark matter.  Rather than having two dark matter components, one hot and one cold, warm dark matter is a single dark matter component with a mean velocity that is greater than that of cold dark matter but less than that of hot dark matter.  One then hopes that warm dark matter is hot enough to reduce the small scale power of CDM but still cold enough not to lose all of the small scale power, as in the case of HDM.

\section{Warm Dark Matter Candidates}

Traditional extensions of the standard model of particle physics usually involve supersymmetry or massive neutrinos.  Both of these scenarios produce interesting dark matter candidates.  Depending on the scale of supersymmetry breaking, either gravitinos or neutralinos may be produced with cosmologically interesting abundances.  At present, the two best warm dark matter candidates are gravitinos or right-handed neutrinos.  Here we consider the motivations for these particles.

\subsection{Supersymmetry}
Perhaps the most promising extension to the standard model is supersymmetry, which is a symmetry relating bosons (integral spins) to fermions (half-integer spins).  In supersymmetry, every known particle has a supersymmetric partner which differs in spin by $\frac{1}{2}$.  As an example, the photon (spin 1) has a supersymmetric partner called the photino (spin $\frac{1}{2}$).  Supersymmetry has a number of features, perhaps the most appealing of which is that it stabilizes the low energy sector against radiative corrections that would otherwise drive the Higgs mass up to the Planck scale.  It also seems to be an essential element for string theory, the leading candidate for a theory of everything.
  
The fact that none of the supersymmetric partners have been observed means that supersymmetry must currently be broken.  In the old days, it was thought that supersymmetry was broken at low energy scales.  In this case the gravitino, which is the spin $\frac{3}{2}$ partner of the graviton (spin 2), has a mass \cite{Pagels}
\begin{equation}
  m_{3/2} = \sqrt{8\pi G/6}F \stackrel{<}{\sim} \mbox{keV},
\end{equation}
where $F$ is the Goldstino decay constant, which is the expectation value of the $F$ component of some hidden sector field responsible for breaking supersymmetry.   Since it is reasonable to expect that the threshold energies for the SUSY particles are comparable to the symmetry breaking scale $F$, the gravitino is expected to be the lightest supersymmetric particle or LSP.  Pagels and Primack proposed the $\pm \frac{1}{2}$ helicity component of the spin $\frac{3}{2}$ gravitino as a dark matter candidate.  This helicity component, which couples much more strongly to matter than the $\pm \frac{3}{2}$ component \cite{Fayet1, Fayet2, Fayet3}, is able to remain in thermal equilibrium until about the electroweak phase transition, when it decouples and freezes out, leaving a relic abundance of gravitinos.

Low energy SUSY breaking is an idea that has received some renewed interest lately in the models of Dine et. al  \cite{Dine1, Dine2, Dine3, Dvali} and as a result of the CDF $\gamma \gamma e^{+}e^{-}$ event \cite{CDF} (although one event is hardly compelling, much less statistically significant).  In these models, it is assumed that there is a hidden sector which consists of a new set of fields and interactions responsible for breaking supersymmetry.  If these fields carry the quantum numbers of the standard model such as color, electric charge, etc., they will couple to the supersymmetric particles through gauge loops.  This is known as gauge mediated supersymmetry breaking.  As was the case in the model of Primack and Pagels, the gravitino is again the LSP with a mass
\begin{equation}
  m_{3/2} = 2.5\left(\frac{F}{(100\mbox{TeV})^{2}}\right)\mbox{eV}
\end{equation}
It is interesting to note that in the context of gauge-mediated SUSY breaking, the interpretation of the CDF $\gamma \gamma e^{+}e^{-}$ event requires $\sqrt{F}\stackrel{<}{\sim}10^{3}$TeV \cite{cdf1,cdf2,cdf3,cdf4}, which means that the gravitino mass $m_{3/2}$ should be less than about 250eV.  Since the gravitino is the lightest supersymmetric particle, all other supersymmetric particles will decay into the gravitino in a cosmologically short time scale \cite{Dimopoulos}.  This raises the rather interesting possibility of having two populations of gravitinos, one of which is produced by freezeout (discussed below), and another which is produced non-thermally by sparticle decays \cite{BM, BMY}.

\subsection{Neutrinos}
In the standard model, fermion Dirac masses arise from terms in the Lagrangian
\begin{equation}
  {\cal L} = -y\bar{f}_{L}\Phi_{H}f_{R},
\end{equation}
where $y$ is the Yukawa coupling of the Higgs field $\Phi_{H}$ to the fermion field $f$, which has been decomposed into left ($L$) and right ($R$) handed components.  When the Higgs field acquires a vacuum expectation value $\sigma/\sqrt{2}$, ${\cal L}_{m} = (y\sigma/\sqrt{2})\bar{f}_{L}f_{R}$, giving the fermion a mass $m_{f}=y\sigma/\sqrt{2}$.  In the electroweak model, though, there are no right-handed neutrinos and so the neutrinos are massless.

Massive neutrinos have many advantages, however.  They seem to be the best explanation for the solar neutrino puzzle, in which the observed Boron neutrino flux is only about 1/3 of what is expected and the observed Beryllium neutrino flux is only about 60\% of what is expected.\footnote{For a review, see  \cite{Bahcall}}  Furthermore, models of structure formation in which some of the dark matter consists of neutrinos with a total mass of about 5eV do quite well when compared to the observational data \cite{Primack}.

Although there are no right-handed neutrinos in the electroweak theory, right handed neutrinos are possible in extensions of the standard model.  Typically in such models, the mass term in the Lagrangian is \cite{Kayser}
\begin{equation}
  {\cal L} = \bar{\nu}{\cal M}_{\nu}\nu,
\end{equation}
where
\begin{equation}
  {\cal M}_{\nu} = \left(\begin{array}{ll}0 &m_{D}\\
  m_{D} &M_{R}\end{array} \right).
\end{equation}
is the mass matrix of the neutrinos, $m_{D}$ is the Dirac ``mass" of the standard model neutrino interaction eigenstates and $M_{R}$ is the ``mass" of the right-handed neutrino interaction eigenstates.  Note that because the physical states are a mixture of the interaction eigenstates, the masses of the physical states are not necessarily equal to $m_{D}$ and $M_{R}$.  Actually, the physical masses are the eigenvalues of the mass matrix ${\cal M}_{\nu}$.  In the case that $M_{R} >> m_{D}$, which occurs in many models, the physical neutrino masses are just $m_{\nu}\sim m^{2}_{D}/M_{R}$ and $m_{N}\sim M_{R}$.  This is known as the seesaw mechanism \cite{seesaw1,seesaw2,seesaw3,seesaw4}.  It seems reasonable to expect that the Dirac mass terms $m_{D}$ are of the same order as the other low energy Dirac mass terms, $m_{D}\stackrel{<}{\sim}$GeV; then, if $M_{R}$ (and therefore $m_{N}$) is large, $m_{\nu}m_{N}=m^{2}_{D}$, which explains why ordinary neutrinos ($m_{\nu}$) are so light.  Since we have no idea what the values of $m_{D}$ and $M_{R}$ might be, it is not hard to imagine that there might be a suitable combination of $m_{D}$ and $M_{R}$ which could give masses $m_{\nu}$ in the range of 1-10eV, just right for hot dark matter.  Another interesting consequence of massive neutrinos is that of oscillations.  Since the mass eigenstates are mixtures of the interaction eigenstates, it is possible for interactions to change one kind of neutrino into another, thereby explaining the solar neutrino puzzle.

In addition to dark matter, which consists of eV mass neutrinos, several authors have considered the idea of a population of heavier sterile neutrinos.  Using the seesaw model, Dodelson and Widrow \cite{DodelsonWidrow} considered the possibility that neutrino oscillations could produce a population of sterile neutrinos, but not at a rate fast enough for the sterile neutrinos to reach thermal equilibrium.  In this scenario, the left-handed neutrino masses are tiny:  $m_{\nu L}\sim 0.22$eV, much too small to be cosmologically interesting.

Another possibility is the model of Malaney, Starkman, and Widrow \cite{Malaney}.
Typically when one considers right-handed neutrinos, it is assumed that all of the right-handed neutrino fields are very massive ($M\sim10^{14}$GeV).  Malaney et al. relaxed this assumption and supposed that one of the right handed neutrinos has a much lower mass ($M'\sim$ 200eV).  In this model there are light neutrinos ($m\approx 25h^{2}$eV) and a heavier, sterile neutrino ($M'\approx 700h^{2}$eV), which is a mixed dark matter scenario.  A supposed advantage of this model is that all of the dark matter can be placed in the neutrino sector; the drawback is that one has to impose a rather unnatural hierarchy among the masses of the right-handed fields:  $M'\ll M$.

\section{The WDM Distribution Function}
The dark matter distribution function can be written as\footnote{Note that $f$ reduces to the distribution function for hot dark matter (such as a light neutrino) when $\alpha_{w}=(4/11)^{1/3}$ and $\beta_{w}=1$.  When $m_{w}\rightarrow\infty$ and $\beta_{w}\rightarrow0$, $f$ describes cold dark matter.} \cite{Colombi}
\begin{equation}
  f = \frac{2}{h^{3}} \frac{\beta_{w}}{e^{p_{0}/\alpha_{w} k_{B}T_{\gamma}} + 1}, \label{eq:distribution}
\end{equation}
where $p_{0}=\sqrt{p^{2}+m_{w}^{2}}$ and $m_{w}$ is the mass of the warm dark matter particle.  The distribution function has three free parameters:  $m_{w}, \alpha_{w}$ and $\beta_{w}$.  Fixing the warm-dark matter density gives one constraint:
\begin{equation}
\Omega_{w}h^{2}=\beta_{w}\frac{\alpha_{w}^{3}}{(4/11)}\left(\frac{m_{w}}{93\mbox{eV}}\right),\label{eq:omegawarm}
\end{equation}
leaving two free parameters, $\alpha_{w}$ and $\beta_{w}$.

The parameter $\alpha_{w}$ depends on the temperature $T_{w}$ of the dark matter particle when it decouples from the radiation field:
\begin{equation}
\alpha_{w}=\frac{T_{w}}{T_{\gamma}}=\left(\frac{4}{11}\right)^{1/3}\left(\frac{10.75}{g_{\ast}(T_{D})}\right)^{1/3},
\end{equation}
where 10.75 is the number of degrees of freedom at ordinary neutrino-decoupling and $g_{\ast}(T_{D})$ is the number of degrees of freedom at WDM decoupling.  In the standard model of particle physics, $g_{\ast}(T_{D})$ has the approximate temperature dependence \cite{KT}
\begin{equation}
  g_{\ast}(T_{D}) \simeq \left\{\begin{array}{ll}3.36, &T \ll \mbox{MeV}\\
  10.75, &\mbox{MeV} \stackrel{<}{\sim} T \stackrel{<}{\sim} 100\mbox{MeV}\\
  61, &200\mbox{MeV} \stackrel{<}{\sim} T \stackrel{<}{\sim} 300\mbox{GeV}\\
  106.75, &T \stackrel{>}{\sim} 300\mbox{GeV}
\end{array} \right .
\end{equation}
Let us consider some examples.   Pagels and Primack \cite{Pagels} assumed that $\omega_{T}$, the threshold for the production of supersymmetric particles was of the order of the weak interaction scale, i.e., $\omega_{T}\sim$100 GeV or larger.  In that case, the gravitinos could remain in equilibrium down to temperatures $\approx \omega_{T}$, leading them to expect that $g_{\ast}(T_{D})\sim 100$.  In the model of Dodelson and Widrow \cite{DodelsonWidrow}, for $m_{w}<100$eV, the right-handed neutrinos are produced at temperatures of 10's of MeV, and so $g_{\ast}(T_{D})=10.75$.  

The parameter $\beta_{w}$ depends on how the warm dark matter particles are produced.  Particles that are simply thermal relics, and whose relic abundance just depends on when they freeze out, like hot dark matter, have $\beta_{w}=1$.  A particle which is produced out of thermal equilibrium, however, will typically have $\beta_{w}<1$.  In the model of Dodelson and Widrow, for example, the parameters of the neutrino mass matrix lead to $\beta_{w}<1$.  

Colombi et al. \cite{Colombi} found that a warm dark matter particle with a mass $m_{w}$ and a temperature $T_{w}$, which has a mass to temperature ratio $m_{w}/T_{w}=46\mbox{eV}/T_{\nu}$, where $T_{\nu}=1.9$K, satisfies several observational tests in the linear regime.  These tests included excess power on $25h^{-1}$ scales, bulk velocities on scales of $40h^{-1}-60h^{-1}$Mpc, and normalization to the COBE data.  On the other hand, non-linear simulations suggest that although WDM does a good job of reproducing the observed structure, it also produces too much small-scale power.  WDM may also have trouble forming galaxies early enough.  And recently, Pierpaoli et al. \cite{Pierpaoli} found that warm dark matter models with light gravitinos (100eV $<m_{g}<$ 1keV) have problems producing the required HI abundances within high redshift damped Ly-$\alpha$ systems and sufficient abundances of local galaxy clusters.  It seems reasonable to expect that warm dark matter models consisting of sterile or right-handed neutrinos will have the same difficulties.  Here we propose another test of warm dark matter, one that combines large scale structure statistics and upcoming microwave background anisotropy experiments.  Since the two leading warm dark matter candidates are gravitinos and neutrinos, let us consider each of those in turn.

\section{Gravitinos}
If supersymmetry is broken at low energy scales, the gravitino is typically the lightest supersymmetric particle.  In the model of Pagels and Primack \cite{Pagels}, it was assumed that the threshold for SUSY particle production, above which the gravitino is in thermal equilibrium, was roughly of the order of the weak energy scale.  This still seems reasonable in the context of gauge-mediated SUSY breaking theories.  In this case, $g_{\ast}(T_{D})$ is about 107 or 62, depending on whether the gravitinos freeze out before or after the electroweak phase transition.  Using these assumptions, we can see that if the gravitino is to provide closure density, the value of $m_{3/2}$ will typically be less than about 250eV.  Since the gravitino mass is heavier than that of the hot dark matter ($\sim 2.4$eV), but is much less than that of the typical cold dark matter ($\stackrel{>}{\sim}$GeV), these gravitinos behave like warm dark matter.  
First, consider the case in which all of the dark matter is in the form of warm gravitinos.  In order to be viable dark matter candidates, the gravitinos should be able to reproduce the observed large scale structure properties.  In linear theory, two useful quantities are $\sigma_{8}$ and the excess power $EP$ \cite{WrightEP}.  The excess power compares the mass fluctuation on $25h^{-1}$Mpc scales to that on $8h^{-1}$Mpc scales, relative to that of standard CDM:
\begin{equation}
  EP = \frac{\sigma(25h^{-1}\mbox{Mpc})/\sigma(8h^{-1}\mbox{Mpc})}{\sigma_{CDM}(25h^{-1}\mbox{Mpc})/\sigma_{CDM}(8h^{-1}\mbox{Mpc})}
\end{equation}
Observationally, $EP=1.3\pm 0.15$, which means there is more power on $25h^{-1}$Mpc scales than would be expected from a standard cold dark matter scenario.  In Table~\ref{tab:lssg} we list the values of $\sigma_{8}$ and the excess power for the two gravitino models, one of which freezes out before the electroweak phase transition, when $g_{\ast}(T_{D})\simeq 107$, and the other which freezes out after the electroweak phase transition, when $g_{\ast}(T_{D})\simeq 62$.  As is evident from this table, neither of these gravitino models does any better than sCDM when compared to the large scale structure data. 

\begin{table}
\begin{center}
\begin{tabular}{|c|c|c|} \hline
 Model			&$\sigma_{8}$	&$EP$ \\ \hline
 sCDM				&1.18			&1	\\ \hline
 $g_{\ast}(T_{D})=107$	&1.17			&1    \\ \hline 
 $g_{\ast}(T_{D})=62$	&1.15			&1.02 \\ \hline
\end{tabular}  
\end{center}
\caption{Large scale structure quantities for pure gravitino warm dark matter models.}\label{tab:lssg}
\end{table}

\begin{figure}
\begin{center}
\begin{picture}(0,0)%
\includegraphics{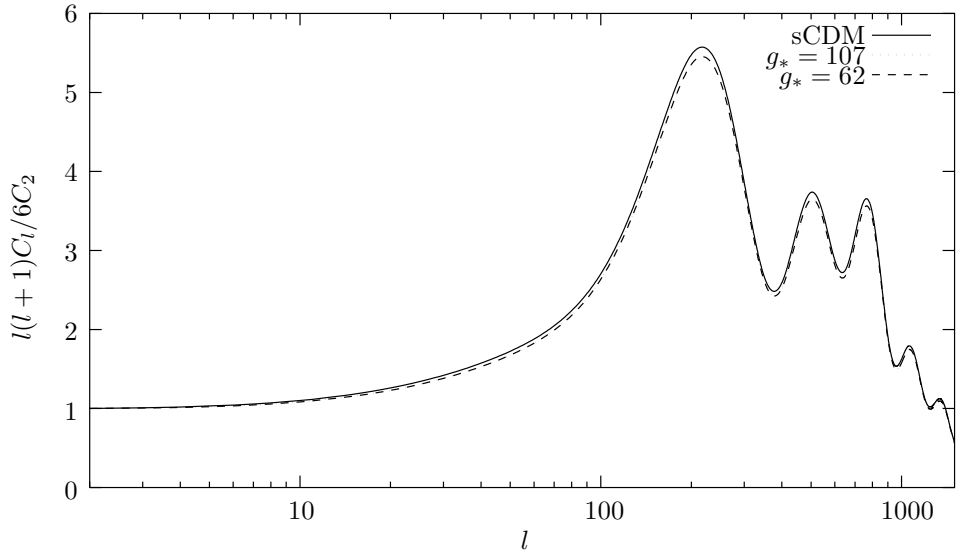}%
\end{picture}%
\setlength{\unitlength}{0.00083300in}%
\begingroup\makeatletter\ifx\SetFigFont\undefined%
\gdef\SetFigFont#1#2#3#4#5{%
  \reset@font\fontsize{#1}{#2pt}%
  \fontfamily{#3}\fontseries{#4}\fontshape{#5}%
  \selectfont}%
\fi\endgroup%
\begin{picture}(5935,3401)(304,-2964)
\put(721,-2627){\makebox(0,0)[rb]{\smash{\SetFigFont{10}{12.0}{\familydefault}{\mddefault}{\updefault}0}}}
\put(721,-2130){\makebox(0,0)[rb]{\smash{\SetFigFont{10}{12.0}{\familydefault}{\mddefault}{\updefault}1}}}
\put(721,-1634){\makebox(0,0)[rb]{\smash{\SetFigFont{10}{12.0}{\familydefault}{\mddefault}{\updefault}2}}}
\put(721,-1137){\makebox(0,0)[rb]{\smash{\SetFigFont{10}{12.0}{\familydefault}{\mddefault}{\updefault}3}}}
\put(721,-640){\makebox(0,0)[rb]{\smash{\SetFigFont{10}{12.0}{\familydefault}{\mddefault}{\updefault}4}}}
\put(721,-144){\makebox(0,0)[rb]{\smash{\SetFigFont{10}{12.0}{\familydefault}{\mddefault}{\updefault}5}}}
\put(721,353){\makebox(0,0)[rb]{\smash{\SetFigFont{10}{12.0}{\familydefault}{\mddefault}{\updefault}6}}}
\put(2116,-2751){\makebox(0,0)[b]{\smash{\SetFigFont{10}{12.0}{\familydefault}{\mddefault}{\updefault}10}}}
\put(4005,-2751){\makebox(0,0)[b]{\smash{\SetFigFont{10}{12.0}{\familydefault}{\mddefault}{\updefault}100}}}
\put(5894,-2751){\makebox(0,0)[b]{\smash{\SetFigFont{10}{12.0}{\familydefault}{\mddefault}{\updefault}1000}}}
\put(425,-1634){\makebox(0,0)[b]{\smash{\SetFigFont{10}{12.0}{\familydefault}{\mddefault}{\updefault}\begin{rotate}{90}$l(l+1)C_{l}/6C_{2}$\end{rotate}}}}
\put(3511,-2937){\makebox(0,0)[b]{\smash{\SetFigFont{10}{12.0}{\familydefault}{\mddefault}{\updefault}$l$}}}
\put(5634,216){\makebox(0,0)[rb]{\smash{\SetFigFont{10}{12.0}{\familydefault}{\mddefault}{\updefault}sCDM}}}
\put(5634, 92){\makebox(0,0)[rb]{\smash{\SetFigFont{10}{12.0}{\familydefault}{\mddefault}{\updefault}$g_{\ast}=107$}}}
\put(5634,-32){\makebox(0,0)[rb]{\smash{\SetFigFont{10}{12.0}{\familydefault}{\mddefault}{\updefault}$g_{\ast}=62$}}}
\end{picture}
\caption[$l(l+1)C_{l}$:  Gravitinos]{The angular power spectrum for the standard cold dark matter model ($\Omega_{b}=0.05, h=0.5, \Omega_{c}=0.95$), and the two gravitino models considered here.} \label{fig:Clwarmg}
\end{center}
\end{figure}

As a curiosity, it is interesting to consider what kind of signature these particles would leave on the microwave background anisotropy.   Figure~\ref{fig:Clwarmg} shows the microwave background anisotropies, which have been calculated using a modified version of the code of Seljak and Zaldarriaga \cite{SZ}, for the two gravitino models as well as the standard cold dark matter model.  Note that the power spectrum is very similar for both models.  It is worth considering whether microwave background satellite experiments could distinguish these models from the standard CDM model.  In the discussion that follows, we shall assume that there are no systematic errors present and that there are no correlations between pixels.  Although this is not strictly true, it is not too far off, since both the MAP and PLANCK satellites should be able to measure the anisotropy to about 1\% \cite{Ned}.  In this case, the uncertainty in the $C_{l}$s is \cite{Jungman}
\begin{equation}
  \sigma_{l}=\sqrt{\frac{2}{(2l+1)f_{sky}}}(C_{l}+w^{-1}e^{l^{2}\sigma_{b}^2}),
\end{equation}
where $\sigma_{b}=7.43x10^{-3}(\theta_{fwhm}/1^{\circ})$, and $f_{sky}$ is the fractional sky coverage of a given experiment.\footnote{Strictly speaking, this formula holds only if there is no correlations between the $C_{l}$s \cite{Knox}.}  This formula is for the case that the whole sky is observed and a fraction $(1-f_{sky})$, such as the Milky Way galaxy, is subtracted.  In calculating the error, we will assume 2/3 sky coverage, which was approximately the sky coverage of COBE.  In practice, the MAP and PLANCK satellite experiments should be able to cover more sky, since they have a much smaller beamsize than COBE, reducing $\sigma_{l}$ \cite{Ned2}.  Notice that for small $l$, the dominant error is due to the $\sqrt{2/(2l+1)/f_{sky}}$ term.  This describes the error due to cosmic variance, i.e., on large angular scales there are only a small number of independent measurements that can be made.  This error becomes relatively unimportant at high $l$, where instead the error is due primarily to the finite resolution of the beam.  Assuming no systematic errors, the MAP satellite, with an angular resolution of $0.2^{\circ}$, has $w^{-1} = 0.8 \times 10^{-15}$ in the 90GHz channel, while the PLANCK satellite, with an angular resolution of $0.17^{\circ}$, has $w^{-1} = 3.3 \times 10^{-18}$ in the 143GHz channel \cite{BJK,MAPw,PLANCKw,PLANCKwhp}.

\begin{figure}
\begin{center}
\begin{picture}(0,0)%
\includegraphics{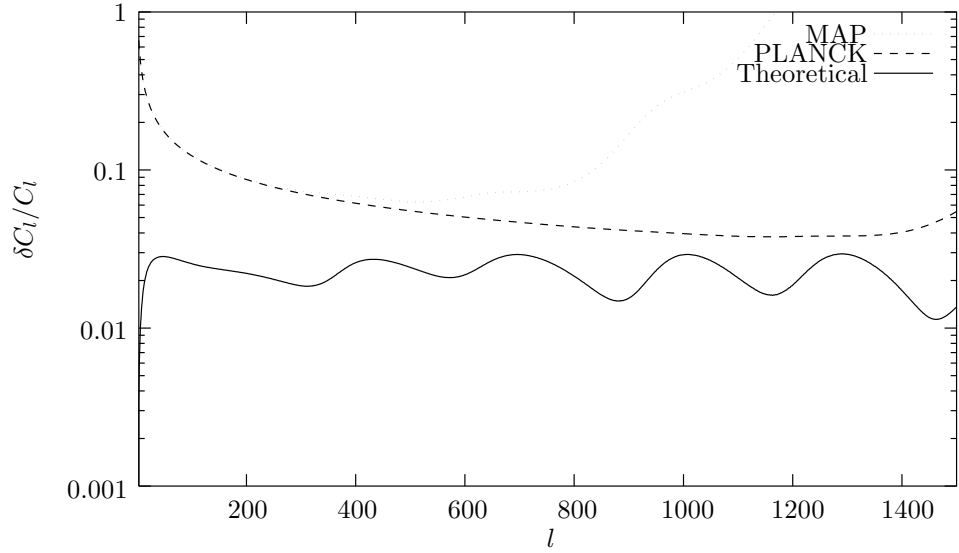}%
\end{picture}%
\setlength{\unitlength}{0.00083300in}%
\begingroup\makeatletter\ifx\SetFigFont\undefined%
\gdef\SetFigFont#1#2#3#4#5{%
  \reset@font\fontsize{#1}{#2pt}%
  \fontfamily{#3}\fontseries{#4}\fontshape{#5}%
  \selectfont}%
\fi\endgroup%
\begin{picture}(5935,3401)(304,-2964)
\put(1017,-2627){\makebox(0,0)[rb]{\smash{\SetFigFont{10}{12.0}{\familydefault}{\mddefault}{\updefault}0.001}}}
\put(1017,-1634){\makebox(0,0)[rb]{\smash{\SetFigFont{10}{12.0}{\familydefault}{\mddefault}{\updefault}0.01}}}
\put(1017,-640){\makebox(0,0)[rb]{\smash{\SetFigFont{10}{12.0}{\familydefault}{\mddefault}{\updefault}0.1}}}
\put(1017,353){\makebox(0,0)[rb]{\smash{\SetFigFont{10}{12.0}{\familydefault}{\mddefault}{\updefault}1}}}
\put(1767,-2751){\makebox(0,0)[b]{\smash{\SetFigFont{10}{12.0}{\familydefault}{\mddefault}{\updefault}200}}}
\put(2453,-2751){\makebox(0,0)[b]{\smash{\SetFigFont{10}{12.0}{\familydefault}{\mddefault}{\updefault}400}}}
\put(3139,-2751){\makebox(0,0)[b]{\smash{\SetFigFont{10}{12.0}{\familydefault}{\mddefault}{\updefault}600}}}
\put(3825,-2751){\makebox(0,0)[b]{\smash{\SetFigFont{10}{12.0}{\familydefault}{\mddefault}{\updefault}800}}}
\put(4512,-2751){\makebox(0,0)[b]{\smash{\SetFigFont{10}{12.0}{\familydefault}{\mddefault}{\updefault}1000}}}
\put(5198,-2751){\makebox(0,0)[b]{\smash{\SetFigFont{10}{12.0}{\familydefault}{\mddefault}{\updefault}1200}}}
\put(5884,-2751){\makebox(0,0)[b]{\smash{\SetFigFont{10}{12.0}{\familydefault}{\mddefault}{\updefault}1400}}}
\put(3659,-2937){\makebox(0,0)[b]{\smash{\SetFigFont{10}{12.0}{\familydefault}{\mddefault}{\updefault}$l$}}}
\put(425,-1137){\makebox(0,0)[b]{\smash{\SetFigFont{10}{12.0}{\familydefault}{\mddefault}{\updefault}\begin{rotate}{90}$\delta C_{l}/C_{l}$\end{rotate}}}}
\put(5634,216){\makebox(0,0)[rb]{\smash{\SetFigFont{10}{12.0}{\familydefault}{\mddefault}{\updefault}MAP}}}
\put(5634, 92){\makebox(0,0)[rb]{\smash{\SetFigFont{10}{12.0}{\familydefault}{\mddefault}{\updefault}PLANCK}}}
\put(5634,-32){\makebox(0,0)[rb]{\smash{\SetFigFont{10}{12.0}{\familydefault}{\mddefault}{\updefault}Theoretical}}}
\end{picture}
\caption[Theoretical and experimental $\delta C_{l}/C_{l}$ for sCDM and pure gravitino WDM which freezes out before the electroweak phase transition]{Theoretical and experimental $\delta C_{l}/C_{l}$ for sCDM and pure gravitino WDM which freezes out before the electroweak phase transition} \label{fig:dw2}
\end{center}
\end{figure}

\begin{figure}
\begin{center}
\begin{picture}(0,0)%
\includegraphics{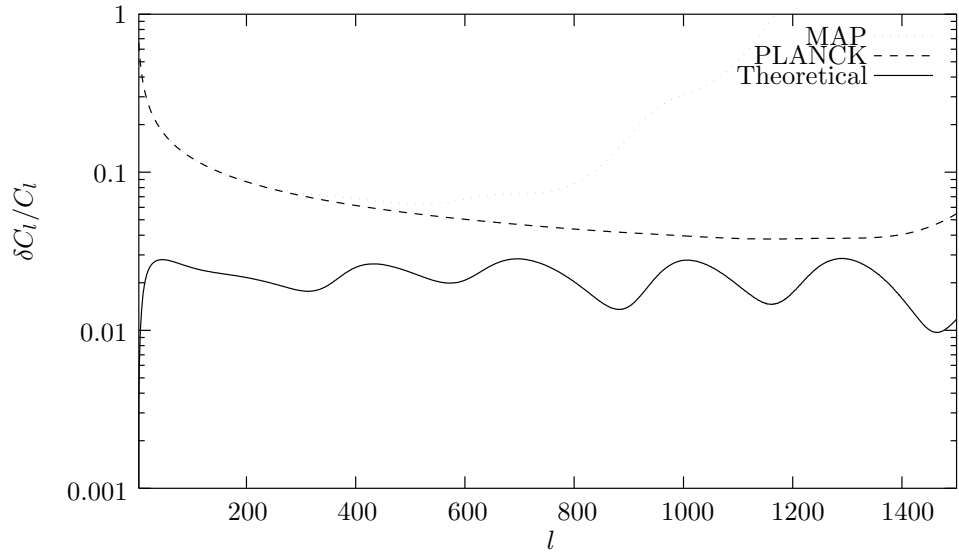}%
\end{picture}%
\setlength{\unitlength}{0.00083300in}%
\begingroup\makeatletter\ifx\SetFigFont\undefined%
\gdef\SetFigFont#1#2#3#4#5{%
  \reset@font\fontsize{#1}{#2pt}%
  \fontfamily{#3}\fontseries{#4}\fontshape{#5}%
  \selectfont}%
\fi\endgroup%
\begin{picture}(5935,3401)(304,-2964)
\put(1017,-2627){\makebox(0,0)[rb]{\smash{\SetFigFont{10}{12.0}{\familydefault}{\mddefault}{\updefault}0.001}}}
\put(1017,-1634){\makebox(0,0)[rb]{\smash{\SetFigFont{10}{12.0}{\familydefault}{\mddefault}{\updefault}0.01}}}
\put(1017,-640){\makebox(0,0)[rb]{\smash{\SetFigFont{10}{12.0}{\familydefault}{\mddefault}{\updefault}0.1}}}
\put(1017,353){\makebox(0,0)[rb]{\smash{\SetFigFont{10}{12.0}{\familydefault}{\mddefault}{\updefault}1}}}
\put(1767,-2751){\makebox(0,0)[b]{\smash{\SetFigFont{10}{12.0}{\familydefault}{\mddefault}{\updefault}200}}}
\put(2453,-2751){\makebox(0,0)[b]{\smash{\SetFigFont{10}{12.0}{\familydefault}{\mddefault}{\updefault}400}}}
\put(3139,-2751){\makebox(0,0)[b]{\smash{\SetFigFont{10}{12.0}{\familydefault}{\mddefault}{\updefault}600}}}
\put(3825,-2751){\makebox(0,0)[b]{\smash{\SetFigFont{10}{12.0}{\familydefault}{\mddefault}{\updefault}800}}}
\put(4512,-2751){\makebox(0,0)[b]{\smash{\SetFigFont{10}{12.0}{\familydefault}{\mddefault}{\updefault}1000}}}
\put(5198,-2751){\makebox(0,0)[b]{\smash{\SetFigFont{10}{12.0}{\familydefault}{\mddefault}{\updefault}1200}}}
\put(5884,-2751){\makebox(0,0)[b]{\smash{\SetFigFont{10}{12.0}{\familydefault}{\mddefault}{\updefault}1400}}}
\put(3659,-2937){\makebox(0,0)[b]{\smash{\SetFigFont{10}{12.0}{\familydefault}{\mddefault}{\updefault}$l$}}}
\put(425,-1137){\makebox(0,0)[b]{\smash{\SetFigFont{10}{12.0}{\familydefault}{\mddefault}{\updefault}\begin{rotate}{90}$\delta C_{l}/C_{l}$\end{rotate}}}}
\put(5634,-32){\makebox(0,0)[rb]{\smash{\SetFigFont{10}{12.0}{\familydefault}{\mddefault}{\updefault}Theoretical}}}
\put(5634, 92){\makebox(0,0)[rb]{\smash{\SetFigFont{10}{12.0}{\familydefault}{\mddefault}{\updefault}PLANCK}}}
\put(5634,216){\makebox(0,0)[rb]{\smash{\SetFigFont{10}{12.0}{\familydefault}{\mddefault}{\updefault}MAP}}}
\end{picture}
\caption[Theoretical and experimental $\delta C_{l}/C_{l}$ for sCDM and pure gravitino WDM which freezes out after the electroweak phase transition]{Theoretical and experimental $\delta C_{l}/C_{l}$ for sCDM and pure gravitino WDM which freezes out after the electroweak phase transition.} \label{fig:dw3}
\end{center}
\end{figure}

\begin{figure}
\begin{center}
\begin{picture}(0,0)%
\includegraphics{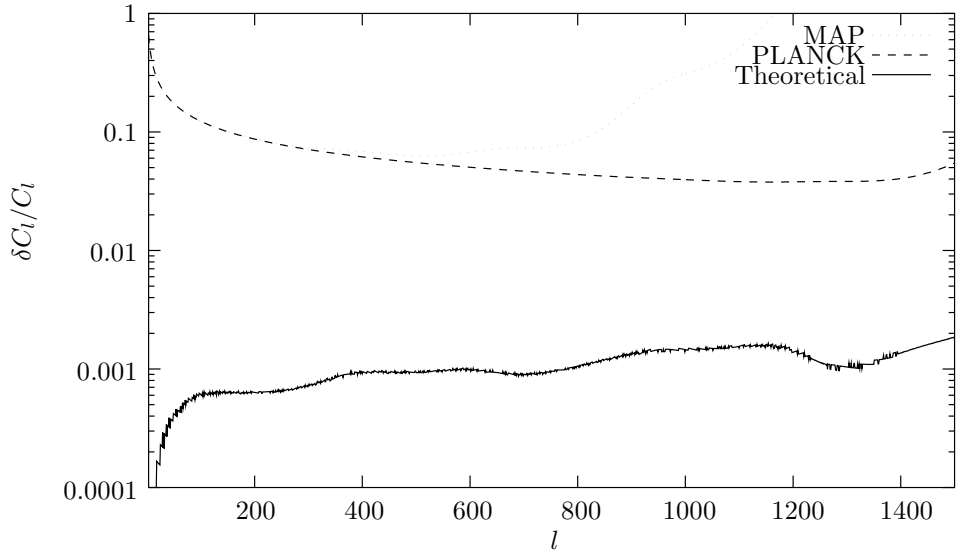}%
\end{picture}%
\setlength{\unitlength}{0.00083300in}%
\begingroup\makeatletter\ifx\SetFigFont\undefined%
\gdef\SetFigFont#1#2#3#4#5{%
  \reset@font\fontsize{#1}{#2pt}%
  \fontfamily{#3}\fontseries{#4}\fontshape{#5}%
  \selectfont}%
\fi\endgroup%
\begin{picture}(5935,3401)(304,-2964)
\put(1091,-2627){\makebox(0,0)[rb]{\smash{\SetFigFont{10}{12.0}{\familydefault}{\mddefault}{\updefault}0.0001}}}
\put(1091,-1882){\makebox(0,0)[rb]{\smash{\SetFigFont{10}{12.0}{\familydefault}{\mddefault}{\updefault}0.001}}}
\put(1091,-1137){\makebox(0,0)[rb]{\smash{\SetFigFont{10}{12.0}{\familydefault}{\mddefault}{\updefault}0.01}}}
\put(1091,-392){\makebox(0,0)[rb]{\smash{\SetFigFont{10}{12.0}{\familydefault}{\mddefault}{\updefault}0.1}}}
\put(1091,353){\makebox(0,0)[rb]{\smash{\SetFigFont{10}{12.0}{\familydefault}{\mddefault}{\updefault}1}}}
\put(1831,-2751){\makebox(0,0)[b]{\smash{\SetFigFont{10}{12.0}{\familydefault}{\mddefault}{\updefault}200}}}
\put(2507,-2751){\makebox(0,0)[b]{\smash{\SetFigFont{10}{12.0}{\familydefault}{\mddefault}{\updefault}400}}}
\put(3184,-2751){\makebox(0,0)[b]{\smash{\SetFigFont{10}{12.0}{\familydefault}{\mddefault}{\updefault}600}}}
\put(3860,-2751){\makebox(0,0)[b]{\smash{\SetFigFont{10}{12.0}{\familydefault}{\mddefault}{\updefault}800}}}
\put(4536,-2751){\makebox(0,0)[b]{\smash{\SetFigFont{10}{12.0}{\familydefault}{\mddefault}{\updefault}1000}}}
\put(5213,-2751){\makebox(0,0)[b]{\smash{\SetFigFont{10}{12.0}{\familydefault}{\mddefault}{\updefault}1200}}}
\put(5889,-2751){\makebox(0,0)[b]{\smash{\SetFigFont{10}{12.0}{\familydefault}{\mddefault}{\updefault}1400}}}
\put(3696,-2937){\makebox(0,0)[b]{\smash{\SetFigFont{10}{12.0}{\familydefault}{\mddefault}{\updefault}$l$}}}
\put(425,-1137){\makebox(0,0)[b]{\smash{\SetFigFont{10}{12.0}{\familydefault}{\mddefault}{\updefault}\begin{rotate}{90}$\delta C_{l}/C_{l}$\end{rotate}}}}
\put(5634,216){\makebox(0,0)[rb]{\smash{\SetFigFont{10}{12.0}{\familydefault}{\mddefault}{\updefault}MAP}}}
\put(5634, 92){\makebox(0,0)[rb]{\smash{\SetFigFont{10}{12.0}{\familydefault}{\mddefault}{\updefault}PLANCK}}}
\put(5634,-32){\makebox(0,0)[rb]{\smash{\SetFigFont{10}{12.0}{\familydefault}{\mddefault}{\updefault}Theoretical}}}
\end{picture}
\caption[Theoretical and experimental $\delta C_{l}/C_{l}$ for the two gravitino models]{Theoretical and experimental $\delta C_{l}/C_{l}$ for the two gravitino models.} \label{fig:dc23}
\end{center}
\end{figure}

To get a rough idea of whether or not the gravitino models are experimentally distinguishable from sCDM, we can compare the experimental sensitivity to the percent difference between the gravitino models and sCDM.  Figures~\ref{fig:dw2} and~\ref{fig:dw3} show $\delta C_{l}/C_{l}$ for the two gravitino models with respect to standard cold dark matter, as well as the precision with which the $C_{l}$'s can be measured by the MAP and PLANCK satellites.  Naively, the difference between the WDM and sCDM is below the threshold of both MAP and PLANCK's sensitivity.  But in the interval $400 \leq l \leq 600$, where the signal to noise ratio is about 1/2, averaging over the 200 multipoles can boost the experimental sensitivity of MAP by a factor of $\sqrt{200} \approx 14$, resulting in a signal to noise ratio of about 7.  Things are even better for PLANCK.  Another cause for optimism is that both experiments will have a larger sky coverage, and PLANCK will probably have a better sensitivity than used in these calculations.  As Figure~\ref{fig:dc23} shows, however, neither satellite could distinguish between these two warm dark matter scenarios.  Equivalently, for a fixed $\Omega_{w}$ and $\beta_{w}$, the microwave background anisotropy is not very sensitive to the mass of the gravitino.

\subsection{Hot and Warm Gravitinos}
Since gravitinos do no better than sCDM in terms of large scale structure, is it still possible for them to comprise the dark matter?  Perhaps.  Since the gravitino is the lightest supersymmetric particle, it can also be produced by the decay of heavier supersymmetric particles.  LEP searches for supersymmetric particles give a lower limit of $\approx 17$GeV \cite{search} for neutral supersymmetric particles.  Because the gravitinos have a mass $m_{3/2}\stackrel{<}{\sim}$keV, the gravitinos that are produced in these decays will be extremely relativistic \cite{BM}.  Therefore it may be possible to have two populations of gravitinos, one of which is a thermal relic that freezes out in the early universe and behaves like warm dark matter, and another that is produced non-thermally by sparticle decays and behaves like hot dark matter.  In the discussion that follows, we will assume that the sparticles that decay into the gravitinos do so very early in the universe, well before galaxies begin forming.  In this scenario, the non-thermal distribution of gravitinos is cold now (just like eV neutrinos, which were once relativistic, are non-relativistic now).  Finally, rather than hazarding a guess about the precise details of the model, we will simply use the fact that, cosmologically speaking, the gravitinos produced by sparticle decays behave just like regular neutrino hot dark matter.  Note that unlike hot dark matter, which may have two hot neutrinos, there is only one hot gravitino in this scenario.

As before, we will consider two types of gravitinos, one of which freezes out before the electroweak phase transition and another that freezes out afterward.  We will also consider two values of $\Omega_{h}$:  0.2 and 0.3.  Therefore, the four warm/hot gravitino models are:
\begin{enumerate}
\item   $g_{\ast}(T_{D})=62$, $\Omega_{hot}=0.2$, $m_{g}=101$eV
\item   $g_{\ast}(T_{D})=107$, $\Omega_{hot}=0.2$, $m_{g}=175$eV
\item   $g_{\ast}(T_{D})=62$, $\Omega_{hot}=0.3$, $m_{g}=87$eV
\item   $g_{\ast}(T_{D})=107$, $\Omega_{hot}=0.3$, $m_{g}=151$eV
\end{enumerate}
In all of these models $h=0.5$ and $\Omega_{b}=0.05$, the standard CDM values.  In Table~\ref{tab:lsshwarmg} we list the values of $\sigma_{8}$ and the excess power for sCDM, the best fit mixed dark matter (MDM) model, which has $\Omega_{\nu}=0.2$ with two massive neutrinos and $\Omega_{c}=0.75$ \cite{Primack}, and the four warm/hot gravitino models.  Unlike the case in which all of the dark matter is a warm gravitino, these four models do quite well in accounting for the observed large scale structure properties, recalling that $\sigma_{8}=0.5-0.8$ and $EP=1.3\pm 0.15$.  Notice that Models 3 and 4, which have $\Omega_{hot}=0.3$, have lower values of $\sigma_{8}$ and more excess power than Models 1 and 2, which have $\Omega_{hot}=0.2$.  This is just what would be expected from adding more hot gravitinos to the mix.

\begin{table}
\begin{center}
\begin{tabular}{|c|c|c|} \hline
 Model	&$\sigma_{8}$	&$EP$	\\ \hline
 sCDM	&1.18		&1	\\ \hline
 MDM	&0.74		&1.31	\\ \hline
 1	&0.8		&1.3	\\ \hline
 2	&0.81		&1.28	\\ \hline 
 3	&0.74		&1.41	\\ \hline
 4	&0.75		&1.39	\\ \hline 
\end{tabular}  
\end{center}
\caption[Large scale structure quantities for the gravitino warm/hot dark matter models]{Large scale structure quantities for the gravitino warm/hot dark matter models.  Also shown are the predictions of standard cold dark matter and a mixed dark matter model with $\Omega_{\nu}=0.2, N_{\nu}=2$.}\label{tab:lsshwarmg}
\end{table}

\begin{figure}
\begin{center}
\begin{picture}(0,0)%
\includegraphics{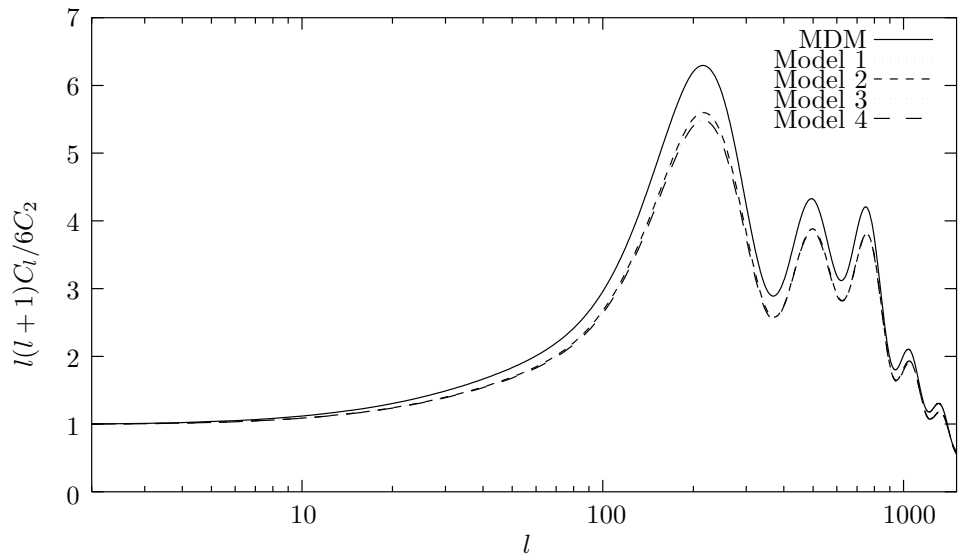}%
\end{picture}%
\setlength{\unitlength}{0.00083300in}%
\begingroup\makeatletter\ifx\SetFigFont\undefined%
\gdef\SetFigFont#1#2#3#4#5{%
  \reset@font\fontsize{#1}{#2pt}%
  \fontfamily{#3}\fontseries{#4}\fontshape{#5}%
  \selectfont}%
\fi\endgroup%
\begin{picture}(5935,3401)(304,-2964)
\put(721,-2627){\makebox(0,0)[rb]{\smash{\SetFigFont{10}{12.0}{\familydefault}{\mddefault}{\updefault}0}}}
\put(721,-2201){\makebox(0,0)[rb]{\smash{\SetFigFont{10}{12.0}{\familydefault}{\mddefault}{\updefault}1}}}
\put(721,-1776){\makebox(0,0)[rb]{\smash{\SetFigFont{10}{12.0}{\familydefault}{\mddefault}{\updefault}2}}}
\put(721,-1350){\makebox(0,0)[rb]{\smash{\SetFigFont{10}{12.0}{\familydefault}{\mddefault}{\updefault}3}}}
\put(721,-924){\makebox(0,0)[rb]{\smash{\SetFigFont{10}{12.0}{\familydefault}{\mddefault}{\updefault}4}}}
\put(721,-498){\makebox(0,0)[rb]{\smash{\SetFigFont{10}{12.0}{\familydefault}{\mddefault}{\updefault}5}}}
\put(721,-73){\makebox(0,0)[rb]{\smash{\SetFigFont{10}{12.0}{\familydefault}{\mddefault}{\updefault}6}}}
\put(721,353){\makebox(0,0)[rb]{\smash{\SetFigFont{10}{12.0}{\familydefault}{\mddefault}{\updefault}7}}}
\put(2116,-2751){\makebox(0,0)[b]{\smash{\SetFigFont{10}{12.0}{\familydefault}{\mddefault}{\updefault}10}}}
\put(4005,-2751){\makebox(0,0)[b]{\smash{\SetFigFont{10}{12.0}{\familydefault}{\mddefault}{\updefault}100}}}
\put(5894,-2751){\makebox(0,0)[b]{\smash{\SetFigFont{10}{12.0}{\familydefault}{\mddefault}{\updefault}1000}}}
\put(425,-1776){\makebox(0,0)[b]{\smash{\SetFigFont{10}{12.0}{\familydefault}{\mddefault}{\updefault}\begin{rotate}{90}$l(l+1)C_{l}/6C_{2}$\end{rotate}}}}
\put(3511,-2937){\makebox(0,0)[b]{\smash{\SetFigFont{10}{12.0}{\familydefault}{\mddefault}{\updefault}$l$}}}
\put(5634,216){\makebox(0,0)[rb]{\smash{\SetFigFont{10}{12.0}{\familydefault}{\mddefault}{\updefault}MDM}}}
\put(5634, 92){\makebox(0,0)[rb]{\smash{\SetFigFont{10}{12.0}{\familydefault}{\mddefault}{\updefault}Model 1}}}
\put(5634,-32){\makebox(0,0)[rb]{\smash{\SetFigFont{10}{12.0}{\familydefault}{\mddefault}{\updefault}Model 2}}}
\put(5634,-156){\makebox(0,0)[rb]{\smash{\SetFigFont{10}{12.0}{\familydefault}{\mddefault}{\updefault}Model 3}}}
\put(5634,-280){\makebox(0,0)[rb]{\smash{\SetFigFont{10}{12.0}{\familydefault}{\mddefault}{\updefault}Model 4}}}
\end{picture}
\caption[$l(l+1)C_{l}$:  MDM and warm/hot gravitino models]{The angular power spectra for an MDM model and the warm/hot gravitino models.  At the top is the MDM model with $\Omega_{\nu}=0.2, N_{\nu}=2$.  Just below it are models 1 and 2 (which are identical), followed by models 3 and 4 (which are also identical).} \label{fig:Clwarmhotgnu}
\end{center}
\end{figure}

\begin{figure}
\begin{center}
\begin{picture}(0,0)%
\includegraphics{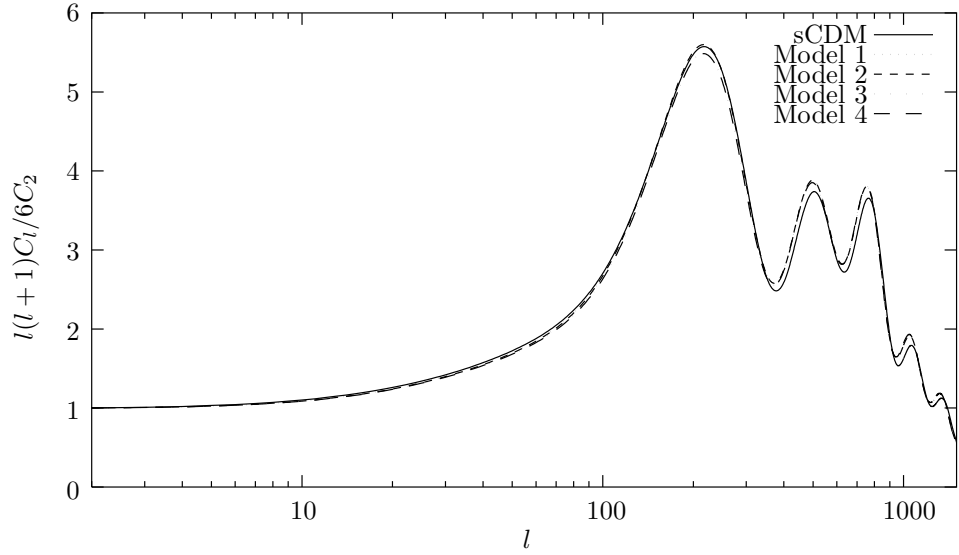}%
\end{picture}%
\setlength{\unitlength}{0.00083300in}%
\begingroup\makeatletter\ifx\SetFigFont\undefined%
\gdef\SetFigFont#1#2#3#4#5{%
  \reset@font\fontsize{#1}{#2pt}%
  \fontfamily{#3}\fontseries{#4}\fontshape{#5}%
  \selectfont}%
\fi\endgroup%
\begin{picture}(5935,3401)(304,-2964)
\put(721,-2627){\makebox(0,0)[rb]{\smash{\SetFigFont{10}{12.0}{\familydefault}{\mddefault}{\updefault}0}}}
\put(721,-2130){\makebox(0,0)[rb]{\smash{\SetFigFont{10}{12.0}{\familydefault}{\mddefault}{\updefault}1}}}
\put(721,-1634){\makebox(0,0)[rb]{\smash{\SetFigFont{10}{12.0}{\familydefault}{\mddefault}{\updefault}2}}}
\put(721,-1137){\makebox(0,0)[rb]{\smash{\SetFigFont{10}{12.0}{\familydefault}{\mddefault}{\updefault}3}}}
\put(721,-640){\makebox(0,0)[rb]{\smash{\SetFigFont{10}{12.0}{\familydefault}{\mddefault}{\updefault}4}}}
\put(721,-144){\makebox(0,0)[rb]{\smash{\SetFigFont{10}{12.0}{\familydefault}{\mddefault}{\updefault}5}}}
\put(721,353){\makebox(0,0)[rb]{\smash{\SetFigFont{10}{12.0}{\familydefault}{\mddefault}{\updefault}6}}}
\put(2116,-2751){\makebox(0,0)[b]{\smash{\SetFigFont{10}{12.0}{\familydefault}{\mddefault}{\updefault}10}}}
\put(4005,-2751){\makebox(0,0)[b]{\smash{\SetFigFont{10}{12.0}{\familydefault}{\mddefault}{\updefault}100}}}
\put(5894,-2751){\makebox(0,0)[b]{\smash{\SetFigFont{10}{12.0}{\familydefault}{\mddefault}{\updefault}1000}}}
\put(425,-1634){\makebox(0,0)[b]{\smash{\SetFigFont{10}{12.0}{\familydefault}{\mddefault}{\updefault}\begin{rotate}{90}$l(l+1)C_{l}/6C_{2}$\end{rotate}}}}
\put(3511,-2937){\makebox(0,0)[b]{\smash{\SetFigFont{10}{12.0}{\familydefault}{\mddefault}{\updefault}$l$}}}
\put(5634,216){\makebox(0,0)[rb]{\smash{\SetFigFont{10}{12.0}{\familydefault}{\mddefault}{\updefault}sCDM}}}
\put(5634, 92){\makebox(0,0)[rb]{\smash{\SetFigFont{10}{12.0}{\familydefault}{\mddefault}{\updefault}Model 1}}}
\put(5634,-280){\makebox(0,0)[rb]{\smash{\SetFigFont{10}{12.0}{\familydefault}{\mddefault}{\updefault}Model 4}}}
\put(5634,-156){\makebox(0,0)[rb]{\smash{\SetFigFont{10}{12.0}{\familydefault}{\mddefault}{\updefault}Model 3}}}
\put(5634,-32){\makebox(0,0)[rb]{\smash{\SetFigFont{10}{12.0}{\familydefault}{\mddefault}{\updefault}Model 2}}}
\end{picture}
\caption[$l(l+1)C_{l}$:  sCDM and warm/hot gravitino models]{The angular power spectra for sMDM and the warm/hot gravitino models.} \label{fig:Clwarmhotgc}
\end{center}
\end{figure}

Now consider the microwave background anisotropy.  Figure~\ref{fig:Clwarmhotgnu} shows the angular power spectra of MDM with $\Omega_{\nu}=0.2$, $N_{\nu}=2$, $\Omega_{b}=0.05$ and $h=0.5$, along with the warm/hot gravitino models. Figure~\ref{fig:Clwarmhotgc} shows the angular power spectra of sCDM and the warm/hot gravitino models.  From these figures, we can see that Models 1 and 2 are indistinguishable, as are Models 3 and 4.  This confirms the results of the last section:  the anisotropy is rather insensitive to $m_{w}$; what is more important in these scenarios is $\Omega_{h}$.  Notice that although all four of the warm/hot gravitino models have a large scale structure which is almost exactly like that of MDM, the microwave background anisotropy of these models is very similar to sCDM but easily distinguishable from MDM.

\begin{figure}
\begin{center}
\begin{picture}(0,0)%
\includegraphics{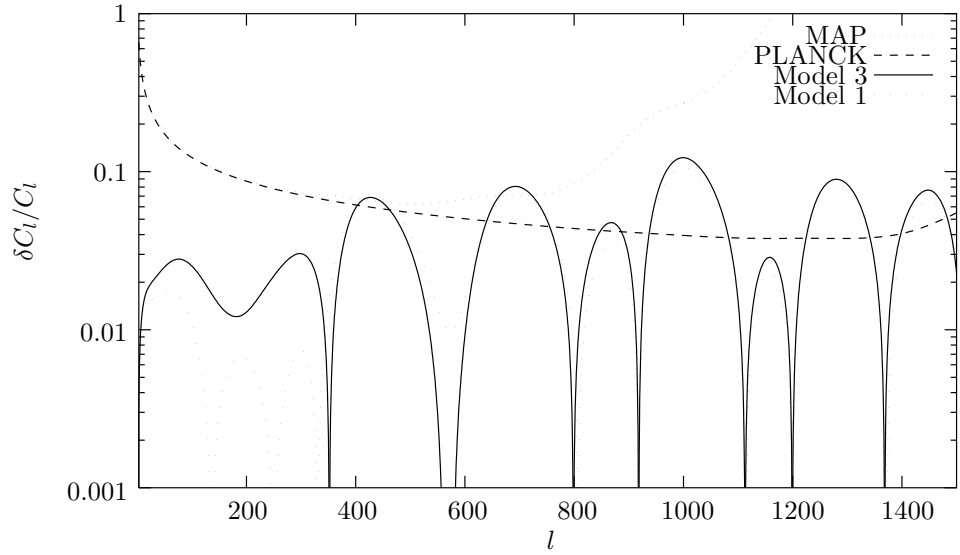}%
\end{picture}%
\setlength{\unitlength}{0.00083300in}%
\begingroup\makeatletter\ifx\SetFigFont\undefined%
\gdef\SetFigFont#1#2#3#4#5{%
  \reset@font\fontsize{#1}{#2pt}%
  \fontfamily{#3}\fontseries{#4}\fontshape{#5}%
  \selectfont}%
\fi\endgroup%
\begin{picture}(5935,3401)(304,-2964)
\put(1017,-2627){\makebox(0,0)[rb]{\smash{\SetFigFont{10}{12.0}{\familydefault}{\mddefault}{\updefault}0.001}}}
\put(1017,-1634){\makebox(0,0)[rb]{\smash{\SetFigFont{10}{12.0}{\familydefault}{\mddefault}{\updefault}0.01}}}
\put(1017,-640){\makebox(0,0)[rb]{\smash{\SetFigFont{10}{12.0}{\familydefault}{\mddefault}{\updefault}0.1}}}
\put(1017,353){\makebox(0,0)[rb]{\smash{\SetFigFont{10}{12.0}{\familydefault}{\mddefault}{\updefault}1}}}
\put(1767,-2751){\makebox(0,0)[b]{\smash{\SetFigFont{10}{12.0}{\familydefault}{\mddefault}{\updefault}200}}}
\put(2453,-2751){\makebox(0,0)[b]{\smash{\SetFigFont{10}{12.0}{\familydefault}{\mddefault}{\updefault}400}}}
\put(3139,-2751){\makebox(0,0)[b]{\smash{\SetFigFont{10}{12.0}{\familydefault}{\mddefault}{\updefault}600}}}
\put(3825,-2751){\makebox(0,0)[b]{\smash{\SetFigFont{10}{12.0}{\familydefault}{\mddefault}{\updefault}800}}}
\put(4512,-2751){\makebox(0,0)[b]{\smash{\SetFigFont{10}{12.0}{\familydefault}{\mddefault}{\updefault}1000}}}
\put(5198,-2751){\makebox(0,0)[b]{\smash{\SetFigFont{10}{12.0}{\familydefault}{\mddefault}{\updefault}1200}}}
\put(5884,-2751){\makebox(0,0)[b]{\smash{\SetFigFont{10}{12.0}{\familydefault}{\mddefault}{\updefault}1400}}}
\put(3659,-2937){\makebox(0,0)[b]{\smash{\SetFigFont{10}{12.0}{\familydefault}{\mddefault}{\updefault}$l$}}}
\put(425,-1137){\makebox(0,0)[b]{\smash{\SetFigFont{10}{12.0}{\familydefault}{\mddefault}{\updefault}\begin{rotate}{90}$\delta C_{l}/C_{l}$\end{rotate}}}}
\put(5634,216){\makebox(0,0)[rb]{\smash{\SetFigFont{10}{12.0}{\familydefault}{\mddefault}{\updefault}MAP}}}
\put(5634, 92){\makebox(0,0)[rb]{\smash{\SetFigFont{10}{12.0}{\familydefault}{\mddefault}{\updefault}PLANCK}}}
\put(5634,-32){\makebox(0,0)[rb]{\smash{\SetFigFont{10}{12.0}{\familydefault}{\mddefault}{\updefault}Model 3}}}
\put(5634,-156){\makebox(0,0)[rb]{\smash{\SetFigFont{10}{12.0}{\familydefault}{\mddefault}{\updefault}Model 1}}}
\end{picture}
\caption[Theoretical and experimental $\delta C_{l}/C_{l}$ for sCDM and the warm/hot gravitino modelsl $\delta C_{l}/C_{l}$ for the MAP and PLANCK  experiments]{Theoretical $\delta C_{l}/C_{l}$ for Model 1 and Model 3 with respect to sCDM, as well as the experimental $\delta C_{l}/C_{l}$ for the MAP and PLANCK  experiments.} \label{fig:dclwhgcdm}
\end{center}
\end{figure}

\begin{figure}
\begin{center}
\begin{picture}(0,0)%
\includegraphics{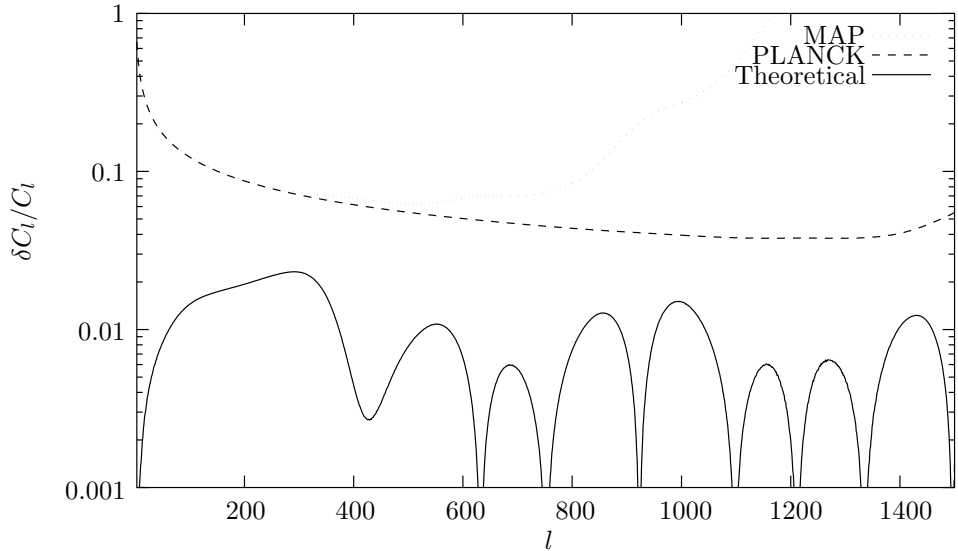}%
\end{picture}%
\setlength{\unitlength}{0.00083300in}%
\begingroup\makeatletter\ifx\SetFigFont\undefined%
\gdef\SetFigFont#1#2#3#4#5{%
  \reset@font\fontsize{#1}{#2pt}%
  \fontfamily{#3}\fontseries{#4}\fontshape{#5}%
  \selectfont}%
\fi\endgroup%
\begin{picture}(5935,3401)(304,-2964)
\put(1017,-2627){\makebox(0,0)[rb]{\smash{\SetFigFont{10}{12.0}{\familydefault}{\mddefault}{\updefault}0.001}}}
\put(1017,-1634){\makebox(0,0)[rb]{\smash{\SetFigFont{10}{12.0}{\familydefault}{\mddefault}{\updefault}0.01}}}
\put(1017,-640){\makebox(0,0)[rb]{\smash{\SetFigFont{10}{12.0}{\familydefault}{\mddefault}{\updefault}0.1}}}
\put(1017,353){\makebox(0,0)[rb]{\smash{\SetFigFont{10}{12.0}{\familydefault}{\mddefault}{\updefault}1}}}
\put(1767,-2751){\makebox(0,0)[b]{\smash{\SetFigFont{10}{12.0}{\familydefault}{\mddefault}{\updefault}200}}}
\put(2453,-2751){\makebox(0,0)[b]{\smash{\SetFigFont{10}{12.0}{\familydefault}{\mddefault}{\updefault}400}}}
\put(3139,-2751){\makebox(0,0)[b]{\smash{\SetFigFont{10}{12.0}{\familydefault}{\mddefault}{\updefault}600}}}
\put(3825,-2751){\makebox(0,0)[b]{\smash{\SetFigFont{10}{12.0}{\familydefault}{\mddefault}{\updefault}800}}}
\put(4512,-2751){\makebox(0,0)[b]{\smash{\SetFigFont{10}{12.0}{\familydefault}{\mddefault}{\updefault}1000}}}
\put(5198,-2751){\makebox(0,0)[b]{\smash{\SetFigFont{10}{12.0}{\familydefault}{\mddefault}{\updefault}1200}}}
\put(5884,-2751){\makebox(0,0)[b]{\smash{\SetFigFont{10}{12.0}{\familydefault}{\mddefault}{\updefault}1400}}}
\put(3659,-2937){\makebox(0,0)[b]{\smash{\SetFigFont{10}{12.0}{\familydefault}{\mddefault}{\updefault}$l$}}}
\put(5634,216){\makebox(0,0)[rb]{\smash{\SetFigFont{10}{12.0}{\familydefault}{\mddefault}{\updefault}MAP}}}
\put(5634, 92){\makebox(0,0)[rb]{\smash{\SetFigFont{10}{12.0}{\familydefault}{\mddefault}{\updefault}PLANCK}}}
\put(5634,-32){\makebox(0,0)[rb]{\smash{\SetFigFont{10}{12.0}{\familydefault}{\mddefault}{\updefault}Theoretical}}}
\put(425,-1137){\makebox(0,0)[b]{\smash{\SetFigFont{10}{12.0}{\familydefault}{\mddefault}{\updefault}\begin{rotate}{90}$\delta C_{l}/C_{l}$\end{rotate}}}}
\end{picture}
\caption[Theoretical and experimental $\delta C_{l}/C_{l}$ for two warm/hot gravitino models]{Theoretical $\delta C_{l}/C_{l}$ for Models 1 and 3, as well as the experimental $\delta C_{l}/C_{l}$ for the MAP and PLANCK experiments.} \label{fig:dcl13g}
\end{center}
\end{figure}

We will conclude this section by asking two questions:
\begin{enumerate}
\item  Are the warm/hot gravitino models distinguishable from sCDM?
\item  Are the warm/hot gravitino models with $\Omega_{h}=0.2$ distinguishable from those with $\Omega_{h}=0.3$?
\end{enumerate}
With regard to the first question, Figure~\ref{fig:dclwhgcdm} shows $\delta C_{l}/C_{l}$ for Models 1 and 3 with respect to sCDM, as well as the sensitivity of the MAP and PLANCK satellites.  As can be seen, both MAP and PLANCK should be able to distinguish between a warm/hot gravitino scenario and sCDM, especially after averaging over the multipoles to increase the signal to noise ratio.  With regard to the second question, Figure~\ref{fig:dcl13g} shows $\delta C_{l}/C_{l}$ for Models 1 and 3.  Averaging over the multipoles should allow PLANCK to distinguish between these scenarios, especially at high $l$.  MAP, however, will probably have considerably more difficulty, since on average the signal to noise from $l=400$ to $l=600$ is less than 1\%.

\section{Right Handed Neutrinos}
Several authors have considered the idea that the dark matter may consist of a population of sterile, right-handed neutrinos.  Dodelson and Widrow \cite{DodelsonWidrow}, for example, investigated the possibility that neutrino oscillations could produce a population of sterile neutrinos, although not fast enough for the sterile neutrinos to reach thermal equilibrium.   In this scenario, the peak production occurs when the temperature is of the order of 10's of MeV, in which case $g_{\ast}=10.75$.  In later work, Colombi et al. \cite{Colombi} showed that a sterile neutrino produced in this way with a mass $m_{w}\sim46$eV does better than sCDM in reproducing large scale structure.  As can be seen in Table~\ref{tab:lss46}, their model has both a more reasonable value of $\sigma_{8}$ and more excess power than either sCDM or the gravitino model in which all of the gravitinos are warm.  But there is a hint in $\sigma_{8}$ that their model still produces too much small scale power, a fact that has been confirmed in non-linear simulations \cite{Colombi}.

\begin{table}
\begin{center}
\begin{tabular}{|c|c|c|} \hline
 Model			&$\sigma_{8}$ &$EP$ \\ \hline
 sCDM			&1.18         &1    \\ \hline
 $m_{w}=46\mbox{eV}$	&0.92         &1.23 \\ \hline 
\end{tabular}  
\end{center}
\caption{Large scale structure quantities for sCDM and a right-handed neutrino model with $m_{\nu}=46$eV.} \label{tab:lss46}
\end{table}

\begin{figure}
\begin{center}
\begin{picture}(0,0)%
\includegraphics{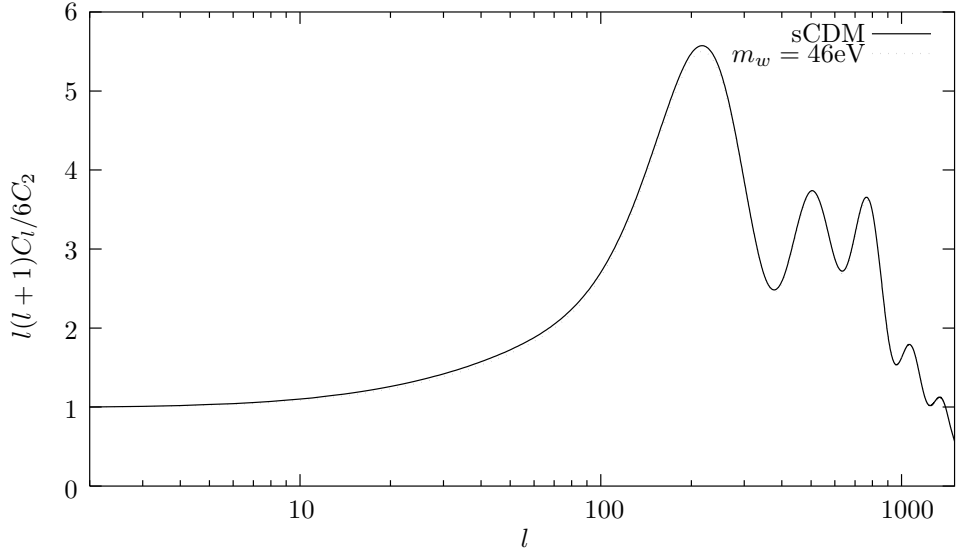}%
\end{picture}%
\setlength{\unitlength}{0.00083300in}%
\begingroup\makeatletter\ifx\SetFigFont\undefined%
\gdef\SetFigFont#1#2#3#4#5{%
  \reset@font\fontsize{#1}{#2pt}%
  \fontfamily{#3}\fontseries{#4}\fontshape{#5}%
  \selectfont}%
\fi\endgroup%
\begin{picture}(5935,3401)(304,-2964)
\put(721,-2627){\makebox(0,0)[rb]{\smash{\SetFigFont{10}{12.0}{\familydefault}{\mddefault}{\updefault}0}}}
\put(721,-2130){\makebox(0,0)[rb]{\smash{\SetFigFont{10}{12.0}{\familydefault}{\mddefault}{\updefault}1}}}
\put(721,-1634){\makebox(0,0)[rb]{\smash{\SetFigFont{10}{12.0}{\familydefault}{\mddefault}{\updefault}2}}}
\put(721,-1137){\makebox(0,0)[rb]{\smash{\SetFigFont{10}{12.0}{\familydefault}{\mddefault}{\updefault}3}}}
\put(721,-640){\makebox(0,0)[rb]{\smash{\SetFigFont{10}{12.0}{\familydefault}{\mddefault}{\updefault}4}}}
\put(721,-144){\makebox(0,0)[rb]{\smash{\SetFigFont{10}{12.0}{\familydefault}{\mddefault}{\updefault}5}}}
\put(721,353){\makebox(0,0)[rb]{\smash{\SetFigFont{10}{12.0}{\familydefault}{\mddefault}{\updefault}6}}}
\put(2116,-2751){\makebox(0,0)[b]{\smash{\SetFigFont{10}{12.0}{\familydefault}{\mddefault}{\updefault}10}}}
\put(4005,-2751){\makebox(0,0)[b]{\smash{\SetFigFont{10}{12.0}{\familydefault}{\mddefault}{\updefault}100}}}
\put(5894,-2751){\makebox(0,0)[b]{\smash{\SetFigFont{10}{12.0}{\familydefault}{\mddefault}{\updefault}1000}}}
\put(425,-1634){\makebox(0,0)[b]{\smash{\SetFigFont{10}{12.0}{\familydefault}{\mddefault}{\updefault}\begin{rotate}{90}$l(l+1)C_{l}/6C_{2}$\end{rotate}}}}
\put(3511,-2937){\makebox(0,0)[b]{\smash{\SetFigFont{10}{12.0}{\familydefault}{\mddefault}{\updefault}$l$}}}
\put(5634,216){\makebox(0,0)[rb]{\smash{\SetFigFont{10}{12.0}{\familydefault}{\mddefault}{\updefault}sCDM}}}
\put(5634, 92){\makebox(0,0)[rb]{\smash{\SetFigFont{10}{12.0}{\familydefault}{\mddefault}{\updefault}$m_{w}=46$eV}}}
\end{picture}
\caption[$l(l+1)C_{l}$:  right-handed neutrino model with $m_{w}=46$eV]{The angular power spectrum for the standard cold dark matter model ($\Omega_{b}=0.05, h=0.5, \Omega_{c}=0.95$), and the best fit warm dark matter model of Colombi et al. with $\Omega_{w}=0.95$.  In this model, the warm dark matter is a sterile right handed neutrino with a mass of 46eV.} \label{fig:Clwarmnu}
\end{center}
\end{figure}

\begin{figure}
\begin{center}
\begin{picture}(0,0)%
\includegraphics{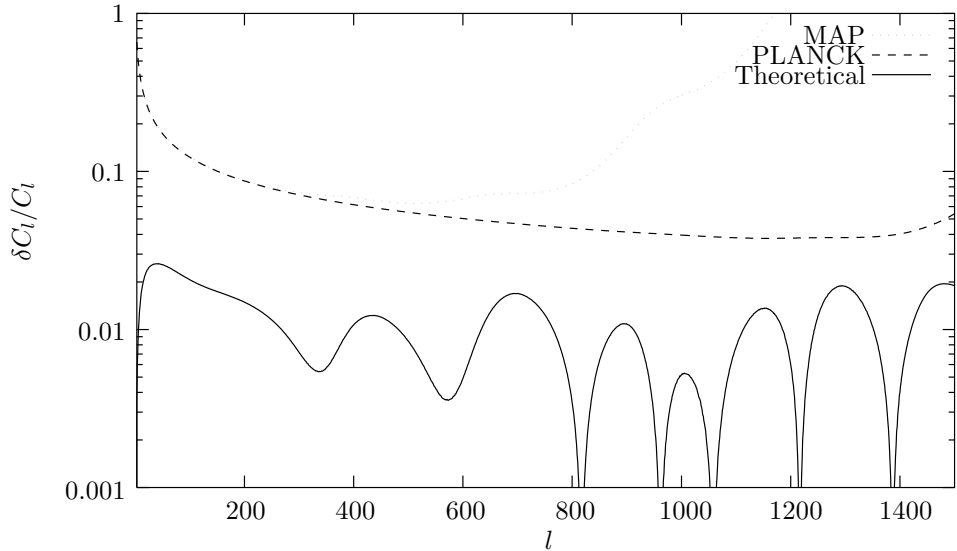}%
\end{picture}%
\setlength{\unitlength}{0.00083300in}%
\begingroup\makeatletter\ifx\SetFigFont\undefined%
\gdef\SetFigFont#1#2#3#4#5{%
  \reset@font\fontsize{#1}{#2pt}%
  \fontfamily{#3}\fontseries{#4}\fontshape{#5}%
  \selectfont}%
\fi\endgroup%
\begin{picture}(5935,3401)(304,-2964)
\put(1017,-2627){\makebox(0,0)[rb]{\smash{\SetFigFont{10}{12.0}{\familydefault}{\mddefault}{\updefault}0.001}}}
\put(1017,-1634){\makebox(0,0)[rb]{\smash{\SetFigFont{10}{12.0}{\familydefault}{\mddefault}{\updefault}0.01}}}
\put(1017,-640){\makebox(0,0)[rb]{\smash{\SetFigFont{10}{12.0}{\familydefault}{\mddefault}{\updefault}0.1}}}
\put(1017,353){\makebox(0,0)[rb]{\smash{\SetFigFont{10}{12.0}{\familydefault}{\mddefault}{\updefault}1}}}
\put(1767,-2751){\makebox(0,0)[b]{\smash{\SetFigFont{10}{12.0}{\familydefault}{\mddefault}{\updefault}200}}}
\put(2453,-2751){\makebox(0,0)[b]{\smash{\SetFigFont{10}{12.0}{\familydefault}{\mddefault}{\updefault}400}}}
\put(3139,-2751){\makebox(0,0)[b]{\smash{\SetFigFont{10}{12.0}{\familydefault}{\mddefault}{\updefault}600}}}
\put(3825,-2751){\makebox(0,0)[b]{\smash{\SetFigFont{10}{12.0}{\familydefault}{\mddefault}{\updefault}800}}}
\put(4512,-2751){\makebox(0,0)[b]{\smash{\SetFigFont{10}{12.0}{\familydefault}{\mddefault}{\updefault}1000}}}
\put(5198,-2751){\makebox(0,0)[b]{\smash{\SetFigFont{10}{12.0}{\familydefault}{\mddefault}{\updefault}1200}}}
\put(5884,-2751){\makebox(0,0)[b]{\smash{\SetFigFont{10}{12.0}{\familydefault}{\mddefault}{\updefault}1400}}}
\put(3659,-2937){\makebox(0,0)[b]{\smash{\SetFigFont{10}{12.0}{\familydefault}{\mddefault}{\updefault}$l$}}}
\put(5634,216){\makebox(0,0)[rb]{\smash{\SetFigFont{10}{12.0}{\familydefault}{\mddefault}{\updefault}MAP}}}
\put(5634, 92){\makebox(0,0)[rb]{\smash{\SetFigFont{10}{12.0}{\familydefault}{\mddefault}{\updefault}PLANCK}}}
\put(5634,-32){\makebox(0,0)[rb]{\smash{\SetFigFont{10}{12.0}{\familydefault}{\mddefault}{\updefault}Theoretical}}}
\put(425,-1137){\makebox(0,0)[b]{\smash{\SetFigFont{10}{12.0}{\familydefault}{\mddefault}{\updefault}\begin{rotate}{90}$\delta C_{l}/C_{l}$\end{rotate}}}}
\end{picture}
\caption[Theoretical and experimental values of $\delta C_{l}/C_{l}$ for sCDM and a warm dark matter model with $m_{w}=46$eV]{Theoretical and experimental values of $\delta C_{l}/C_{l}$ for sCDM and a warm dark matter model with $m_{w}=46$eV.} \label{fig:dClwarmnu}
\end{center}
\end{figure}

What kind of signature would a 46eV sterile neutrino leave on the microwave background?  Figure~\ref{fig:Clwarmnu} shows the microwave background anisotropies for standard CDM and a sterile neutrino model for which $m_{w}=46$eV.  Note that the power spectrum is very similar for both models.  In Figure~\ref{fig:dClwarmnu} we plot $\delta C_{l}/C_{l}$ in the angular power spectrum for a 46eV sterile neutrino with respect to the cold dark matter angular power spectrum, as well as the precision with which the $C_{l}$'s could be measured by the MAP and PLANCK satellites.   As was the case with gravitinos and sCDM, the difference between WDM with $m_{w}=46$eV and sCDM is slightly below the sensitivity of MAP and PLANCK.  But just as before, both satellites should have better sky coverage and PLANCK may have a higher sensitivity.  After averaging over the multipoles, PLANCK should probably be able to distinguish these two models, especially at high $l$.  For MAP, the signal to noise ratio for $400 \leq l \leq 600$ is, on average, roughly 1/10.  Assuming $f_{sky}=2/3$ and $\theta_{fwhm}=0.2^{\circ}$, differentiating sCDM and a 46eV sterile neutrino model will be more challenging for MAP.

\subsection{Hot and Warm Neutrinos}
Since the sterile neutrino WDM model produces too much small scale power, one solution is to add a hot dark matter component.  But when the sterile neutrinos only have masses of the order of 10's of eV, the seesaw mechanism will make the neutrinos so light as to be cosmologically uninteresting.  Malaney et al. \cite{Malaney} suggested that it is possible to have warm and hot neutrinos by imposing a somewhat unnatural hierarchy in the masses of the right-handed neutrino fields.  If, for example, one field has a mass $M'\sim$ 200eV and the others have masses $M\sim10^{14}$GeV, the seesaw mechanism would lead to a population of light neutrinos with $m_{\nu} \approx 25h^{2}$eV and one warm, sterile neutrino with $m_{w} \approx 700h^{2}$eV, i.e., a mixed hot and warm dark matter scenario.

Like the warm gravitino, the sterile neutrino will also freeze out rather early.  In order not to interfere with the nucleosynthesis predictions, it should have decoupled well before nucleosynthesis began.  As in the case for the warm gravitino, then, we will consider the possibilities that the sterile neutrino freezes out either before or after the electroweak phase transition, but before the QCD phase transition.

Here we will not only vary the freeze-out temperature (or $g_{\ast}(T_{D})$), but also the hot neutrino content, $\Omega_{hot}$, and the number, $N_{\nu}$, of hot neutrinos.  The standard hot dark matter scenarios have either $\Omega_{hot}=0.2$ and $N_{\nu}=2$ or $\Omega_{hot}=0.3$ and $N_{\nu}=1$.  We will therefore consider four models:
\begin{enumerate}
\item   $g_{\ast}(T_{D})=62$, $\Omega_{hot}=0.2$, $N_{\nu}=2$, $m_{w}=100$eV
\item   $g_{\ast}(T_{D})=107$, $\Omega_{hot}=0.2$, $N_{\nu}=2$, $m_{w}=174$eV
\item   $g_{\ast}(T_{D})=62$, $\Omega_{hot}=0.3$, $N_{\nu}=1$, $m_{w}=87$eV
\item   $g_{\ast}(T_{D})=107$, $\Omega_{hot}=0.3$, $N_{\nu}=1$, $m_{w}=151$eV
\end{enumerate}
In all of these models, $h=0.5$ and $\Omega_{b}=0.05$, the standard CDM values.  Note that Models 3 and 4 are identical to Models 3 and 4 of the warm/hot gravitino models, since in both cases there is only one massive hot particle and $g_{\ast}(T_{D})$ has the same value at freeze-out.  Therefore, in what follows, we will only discuss the first two models.  In Table~\ref{tab:lsshwarm} we list the values of $\sigma_{8}$ and the excess power for:  standard CDM (sCDM); the best fit mixed dark matter (MDM) model, which has $\Omega_{\nu}=0.2$ with two massive neutrinos and $\Omega_{c}=0.75$; and Models 1 and 2.  Recalling that $EP=1.3\pm 0.15$ and $\sigma_{8}=0.5-0.8$, we can see that Models 1 and 2 agree quite well with the observed large scale structure properties.  
\begin{table}
\begin{center}
\begin{tabular}{|c|c|c|} \hline
 Model	&$\sigma_{8}$	&$EP$	\\ \hline
 sCDM	&1.18		&1	\\ \hline
 MDM	&0.74		&1.31	\\ \hline
 1	&0.72		&1.34	\\ \hline
 2	&0.73		&1.32	\\ \hline 
\end{tabular}
\end{center}  
\caption{Large scale structure quantities for warm/hot right-handed neutrino models.} \label{tab:lsshwarm}
\end{table}

\begin{figure}
\begin{center}
\begin{picture}(0,0)%
\includegraphics{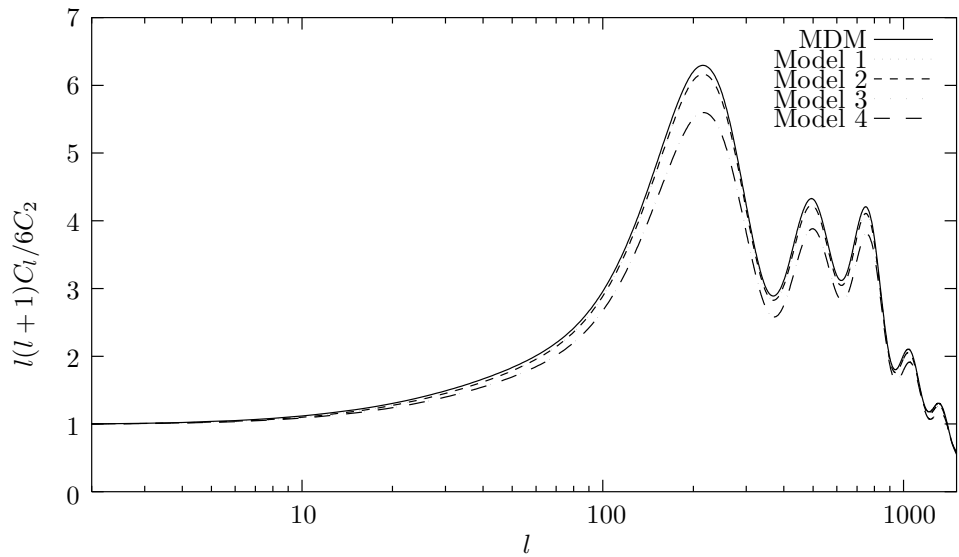}%
\end{picture}%
\setlength{\unitlength}{0.00083300in}%
\begingroup\makeatletter\ifx\SetFigFont\undefined%
\gdef\SetFigFont#1#2#3#4#5{%
  \reset@font\fontsize{#1}{#2pt}%
  \fontfamily{#3}\fontseries{#4}\fontshape{#5}%
  \selectfont}%
\fi\endgroup%
\begin{picture}(5935,3401)(304,-2964)
\put(721,-2627){\makebox(0,0)[rb]{\smash{\SetFigFont{10}{12.0}{\familydefault}{\mddefault}{\updefault}0}}}
\put(721,-2201){\makebox(0,0)[rb]{\smash{\SetFigFont{10}{12.0}{\familydefault}{\mddefault}{\updefault}1}}}
\put(721,-1776){\makebox(0,0)[rb]{\smash{\SetFigFont{10}{12.0}{\familydefault}{\mddefault}{\updefault}2}}}
\put(721,-1350){\makebox(0,0)[rb]{\smash{\SetFigFont{10}{12.0}{\familydefault}{\mddefault}{\updefault}3}}}
\put(721,-924){\makebox(0,0)[rb]{\smash{\SetFigFont{10}{12.0}{\familydefault}{\mddefault}{\updefault}4}}}
\put(721,-498){\makebox(0,0)[rb]{\smash{\SetFigFont{10}{12.0}{\familydefault}{\mddefault}{\updefault}5}}}
\put(721,-73){\makebox(0,0)[rb]{\smash{\SetFigFont{10}{12.0}{\familydefault}{\mddefault}{\updefault}6}}}
\put(721,353){\makebox(0,0)[rb]{\smash{\SetFigFont{10}{12.0}{\familydefault}{\mddefault}{\updefault}7}}}
\put(2116,-2751){\makebox(0,0)[b]{\smash{\SetFigFont{10}{12.0}{\familydefault}{\mddefault}{\updefault}10}}}
\put(4005,-2751){\makebox(0,0)[b]{\smash{\SetFigFont{10}{12.0}{\familydefault}{\mddefault}{\updefault}100}}}
\put(5894,-2751){\makebox(0,0)[b]{\smash{\SetFigFont{10}{12.0}{\familydefault}{\mddefault}{\updefault}1000}}}
\put(425,-1776){\makebox(0,0)[b]{\smash{\SetFigFont{10}{12.0}{\familydefault}{\mddefault}{\updefault}\begin{rotate}{90}$l(l+1)C_{l}/6C_{2}$\end{rotate}}}}
\put(3511,-2937){\makebox(0,0)[b]{\smash{\SetFigFont{10}{12.0}{\familydefault}{\mddefault}{\updefault}$l$}}}
\put(5634,-280){\makebox(0,0)[rb]{\smash{\SetFigFont{10}{12.0}{\familydefault}{\mddefault}{\updefault}Model 4}}}
\put(5634,-156){\makebox(0,0)[rb]{\smash{\SetFigFont{10}{12.0}{\familydefault}{\mddefault}{\updefault}Model 3}}}
\put(5634,-32){\makebox(0,0)[rb]{\smash{\SetFigFont{10}{12.0}{\familydefault}{\mddefault}{\updefault}Model 2}}}
\put(5634, 92){\makebox(0,0)[rb]{\smash{\SetFigFont{10}{12.0}{\familydefault}{\mddefault}{\updefault}Model 1}}}
\put(5634,216){\makebox(0,0)[rb]{\smash{\SetFigFont{10}{12.0}{\familydefault}{\mddefault}{\updefault}MDM}}}
\end{picture}
\caption[$l(l+1)C_{l}$:  MDM and warm/hot right-handed neutrino models]{From top to bottom, the angular power spectrum for MDM, models 1 and 2, and models 3 and 4.} \label{fig:Clwarmhnu}
\end{center}
\end{figure}

\begin{figure}
\begin{center}
\begin{picture}(0,0)%
\includegraphics{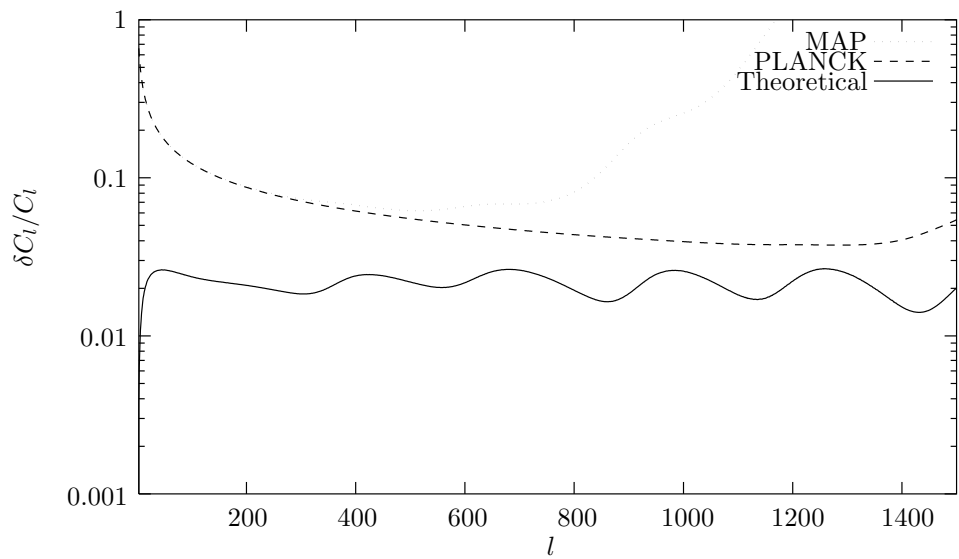}%
\end{picture}%
\setlength{\unitlength}{0.00083300in}%
\begingroup\makeatletter\ifx\SetFigFont\undefined%
\gdef\SetFigFont#1#2#3#4#5{%
  \reset@font\fontsize{#1}{#2pt}%
  \fontfamily{#3}\fontseries{#4}\fontshape{#5}%
  \selectfont}%
\fi\endgroup%
\begin{picture}(5935,3401)(304,-2964)
\put(1017,-2627){\makebox(0,0)[rb]{\smash{\SetFigFont{10}{12.0}{\familydefault}{\mddefault}{\updefault}0.001}}}
\put(1017,-1634){\makebox(0,0)[rb]{\smash{\SetFigFont{10}{12.0}{\familydefault}{\mddefault}{\updefault}0.01}}}
\put(1017,-640){\makebox(0,0)[rb]{\smash{\SetFigFont{10}{12.0}{\familydefault}{\mddefault}{\updefault}0.1}}}
\put(1017,353){\makebox(0,0)[rb]{\smash{\SetFigFont{10}{12.0}{\familydefault}{\mddefault}{\updefault}1}}}
\put(1767,-2751){\makebox(0,0)[b]{\smash{\SetFigFont{10}{12.0}{\familydefault}{\mddefault}{\updefault}200}}}
\put(2453,-2751){\makebox(0,0)[b]{\smash{\SetFigFont{10}{12.0}{\familydefault}{\mddefault}{\updefault}400}}}
\put(3139,-2751){\makebox(0,0)[b]{\smash{\SetFigFont{10}{12.0}{\familydefault}{\mddefault}{\updefault}600}}}
\put(3825,-2751){\makebox(0,0)[b]{\smash{\SetFigFont{10}{12.0}{\familydefault}{\mddefault}{\updefault}800}}}
\put(4512,-2751){\makebox(0,0)[b]{\smash{\SetFigFont{10}{12.0}{\familydefault}{\mddefault}{\updefault}1000}}}
\put(5198,-2751){\makebox(0,0)[b]{\smash{\SetFigFont{10}{12.0}{\familydefault}{\mddefault}{\updefault}1200}}}
\put(5884,-2751){\makebox(0,0)[b]{\smash{\SetFigFont{10}{12.0}{\familydefault}{\mddefault}{\updefault}1400}}}
\put(3659,-2937){\makebox(0,0)[b]{\smash{\SetFigFont{10}{12.0}{\familydefault}{\mddefault}{\updefault}$l$}}}
\put(425,-1137){\makebox(0,0)[b]{\smash{\SetFigFont{10}{12.0}{\familydefault}{\mddefault}{\updefault}\begin{rotate}{90}$\delta C_{l}/C_{l}$\end{rotate}}}}
\put(5634,216){\makebox(0,0)[rb]{\smash{\SetFigFont{10}{12.0}{\familydefault}{\mddefault}{\updefault}MAP}}}
\put(5634, 92){\makebox(0,0)[rb]{\smash{\SetFigFont{10}{12.0}{\familydefault}{\mddefault}{\updefault}PLANCK}}}
\put(5634,-32){\makebox(0,0)[rb]{\smash{\SetFigFont{10}{12.0}{\familydefault}{\mddefault}{\updefault}Theoretical}}}
\end{picture}
\caption[Theoretical and experimental $\delta C_{l}/C_{l}$ for MDM and a warm/hot neutrino model]{Theoretical and experimental $\delta C_{l}/C_{l}$ for MDM and Model 1.} \label{fig:dClwh1}
\end{center}
\end{figure}

Now consider the microwave background anisotropy.  As can be seen in Figure~\ref{fig:Clwarmhnu}, the anisotropy for the warm/hot Models 1 and 2 is pretty close to that of MDM.  This is not surprising, since the warm component is close to being cold whereas the hot component is simply regular hot dark matter.  Since there is no observable difference between Models 1 and 2, i.e., it doesn't seem to matter when the right-handed neutrinos freeze-out, we plot the theoretical $\delta C_{l}/C_{l}$ for Model 1 and MDM as well as the experimental $\delta C_{l}/C_{l}$ for the MAP and PLANCK satellites in Figure~\ref{fig:dClwh1}.   From this figure, it seems reasonable to expect that both MAP and PLANCK can distinguish MDM from the warm/hot neutrino models considered here.

\section{Conclusions}
Warm dark matter is interesting because it offers the possibility of solving the dark matter problem entirely in one new sector of the standard model.     Unfortunately, the simplest warm dark matter models do not appear to work:  like sCDM, they produce too much structure on small scales.  An interesting resolution to this problem is offered by supersymmetry.  If SUSY is broken at relatively low energy scales, it may be possible to have two populations of gravitinos; phenomenologically, this would correspond to a mixture of warm and hot dark matter.

With the exception of the rather unnatural model of Malaney et al. \cite{Malaney}, all of the warm dark matter models have two rather robust features:
\begin{enumerate}
\item  Both WDM and sCDM have a very similar angular power spectrum of microwave background anisotropies.

\item  WDM has essentially the same linear large scale structure properties as MDM.
\end{enumerate}
Both PLANCK and MAP should be able to resolve the difference between sCDM and the gravitino WDM and warm/hot gravitino models.  PLANCK should be able to distinguish between sCDM and a sterile neutrino WDM model, although MAP may have more difficulty.  Nevertheless, these robust features provide a strong test of WDM.  If, for example, large scale structure suggests MDM, but the CMB anisotropies look like those of CDM, this would be a strong signal for WDM.  On the other hand, if the large scale structure looks like MDM and the CMB looks significantly different from CDM, this would be a strong case against WDM.  Therefore we expect that a combination of CMB and large scale structure measurements will confirm or rule out WDM once and for all.

I would like to thank Ned Wright and Roberto Peccei for helpful comments on this work.


\begin{thebibliography}{25}
\bibitem{KT}E. Kolb and M. S. Turner, \begin{em}The Early Universe\end{em} (Addison-Wesley, 1990).
\bibitem{ShafiMDM}Q. Shafi and F. W. Stecker, \begin{em}Phys. Rev. Lett.\end{em} {\bf 53}, 1292 (1984).
\bibitem{DavisMDM}M. Davis, F. Summers and D. Schlegel, \begin{em}Nature\end{em} {\bf 359}, 393 (1992).
\bibitem{TaylorMDM}A. N. Taylor and M. Rowan-Robinson, \begin{em}Nature\end{em} {\bf 359}, 396 (1992).
\bibitem{vanDalenMDM}A. van Dalen and R. K. Schaefer, \begin{em}Astrophys. J.\end{em} {\bf 398}, 33 (1992).
\bibitem{KlypinMDM}A. Klypin, J. Holtzman, J. Primack and E. Reg\"{o}s, \begin{em}Astrophys. J.\end{em} {\bf 416}, 1 (1993).
\bibitem{tiltedCDM}F. Adams, et al.  \begin{em}Phys. Rev.\end{em} {\bf D47}, 426 (1993).
\bibitem{Peebleslambda}P. J. E. Peebles, \begin{em}Astrophys. J.\end{em} {\bf 258}, 415, (1984).
\bibitem{Turnerlambda1}M. S. Turner, G. Steigman and L. Krauss, \begin{em}Phys. Rev. Lett.\end{em} {\bf 52}, 2090 (1984).
\bibitem{Efstathioulambda}G. Efstathiou et al., \begin{em}Nature\end{em}, {\bf 348}, 705 (1990).
\bibitem{Turnerlambda2}M. S. Turner, \begin{em}Phys. Scr.\end{em} {\bf T36}, 167 (1991).
\bibitem{Pagels}H. R. Pagels and J. R. Primack, Phys. Rev. Lett {\bf 48}, 223 (1982).
\bibitem{Fayet1}P. Fayet in \begin{em}Unification of the Fundamental Particle Interactions\end{em}, ed. S. Ferrara, J. Ellis and P. van Nieuwenhuizen (Plenum, New York, 1980).
\bibitem{Fayet2}P. Fayet, \begin{em}Phys. Lett.\end{em} {\bf 70B}, 461 (1977).
\bibitem{Fayet3}P. Fayet, \begin{em}Phys. Lett.\end{em} {\bf 84B}, 421 (1979).
\bibitem{Dine1}M. Dine, A. Nelson, Y. Nir and Y. Shirman, \begin{em}Phys. Rev.\end{em} {\bf D53}, 2658 (1996).
\bibitem{Dine2}M. Dine and A. E. Nelson, \begin{em}Phys. Rev.\end{em} {\bf D48}, 1277 (1993).
\bibitem{Dine3}M. Dine, A. E. Nelson and Y. Shirman, \begin{em}Phys. Rev.\end{em} {\bf D51}, 1362 (1995).
\bibitem{Dvali}G. Dvali, G. F. Giudice and A. Pomarol, \begin{em}Nucl. Phys.\end{em} {\bf B478}, 31 (1996).
\bibitem{CDF}D. Toback (for the CDF Collaboration), ``The Diphoton Missing $E_{t}$ Distribution at CDF," FERMILAB-CONF-96-240-E.
\bibitem{cdf1}S. Dimopoulos, M. Dine, S. Raby and S. Thomas, \begin{em}Phys. Rev. Lett.\end{em} {\bf 76}, 3494 (1996).
\bibitem{cdf2}S. Ambrosanio, G. Kane, G. Kribs, S. Martin and S. Mrenna, \begin{em}Phys. Rev. Lett.\end{em} {\bf 76}, 3498 (1996).
\bibitem{cdf3}S. Dimopoulos, S. Thomas and J. Wells, hep-ph/9604452 (1996).
\bibitem{cdf4}K. S. Babu, C. Kolda and F. Wilczek, hep-ph/9605408 (1996).
\bibitem{Dimopoulos}S. Dimopoulos, G. F. Giudice and A. Pomarol, hep-ph/9607225 (1996).
\bibitem{BM}S. Borgani and A. Masiero, hep-ph/9701417 (1997).
\bibitem{BMY}S. Borgani, A. Masiero, and M. Yamaguchi, hep-ph/9605222 (1996).
\bibitem{Bahcall}J. Bahcall, astro-ph/9702057 (1997).
\bibitem{Primack}J. Primack, to appear in \begin{em}Critical Dialogues in Cosmology\end{em}, ed. N. Turok (World Scientific, 1996).
\bibitem{Kayser}B. Kayser, et al. \begin{em}The Physics of Massive Neutrinos\end{em} (World Scientific, 1989).
\bibitem{seesaw1}M. Gell-Mann, P. Ramond and R. Slansky in \begin{em}Supergravity\end{em}, ed. D. Freedman and P. van Nieuwenhuizen (North Holland, 1979).
\bibitem{seesaw2}T. Yanagida in \begin{em}Proceedings of the Workshop on Unified Theory and Baryon Number in the Universe\end{em}, ed. O. Sawada and A. Sugamoto (KEK, 1979).
\bibitem{seesaw3}R. Mohapatra and G. Senjanovic, \begin{em}Phys. Rev. Lett.\end{em} {\bf 44}, 912 (1980).
\bibitem{seesaw4}R. Mohapatra and G. Senjanovic, \begin{em}Phys. Rev.\end{em} {\bf D23}, 165 (1981).
\bibitem{DodelsonWidrow}S. Dodelson and L. M. Widrow, Phys. Rev. Lett. 72, 17 (1994).
\bibitem{Malaney}R. A. Malaney, G. D. Starkman and L. M. Widrow, Phys. Rev. {\bf D52}, 5480 (1995).
\bibitem{Colombi}S. Colombi, S. Dodelson, and L. M. Widrow, Astrophys. J. 458, 1 (1996).
\bibitem{Pierpaoli}E. Pierpaoli, et al., astro-ph 9709057 (1997).
\bibitem{WrightEP}E. L. Wright et al., \begin{em}Astrophys. J.\end{em} {\bf 396}, L13 (1992).
\bibitem{SZ}U. Seljak and M. Zaldarriaga, Astrophys. J. 469, 437 (1996).
\bibitem{Ned}Ned Wright, private communication.
\bibitem{Jungman}G. Jungman, et. al., Phys. Rev. D54, 1332 (1996).
\bibitem{Knox}L. Knox, astro-ph 9606066(1996).
\bibitem{Ned2}Ned Wright, private communication.
\bibitem{BJK}J. R. Bond, A. H. Jaffe and L. Knox, astro-ph/9708203 (1997).
\bibitem{MAPw}C. L. Bennett et al., MAP experiment home page, http://map.gafc.nasa.gov (1996).
\bibitem{PLANCKw}M. Bersanelli et al., \begin{em}COBRAS/SAMBA, The Phase A Study for an ESA M3 Mission\end{em}, preprint (1996).
\bibitem{PLANCKwhp}M. Bersanelli et al., Planck home page, http://astro.estec.esa.nl/SA-general/Projects/Cobras/cobras.html (1996).
\bibitem{search}J. Ellis, T. Falk, K. Olive, and M. Schmitt, (1996), hep-ph 9610410.
\end{thebibliography}
\end{document}